\documentclass[a4paper,11pt]{article}
\usepackage{jheppub}
\usepackage[T1]{fontenc}

\title{\boldmath The correlation of WGC and Hydrodynamics bound with $R^4$ correction in the charged AdS$_{d+2}$ black brane }

\author[a]{E.Naghd Mezerji,}
\author[a]{J.Sadeghi}

\affiliation[a]{Department of Physics, Faculty of Basic Sciences, University of Mazandaran P. O. Box 47416-95447, Babolsar, Iran}

\emailAdd{e.n.mezerji@stu.umz.ac.ir}
\emailAdd{pouriya@ipm.ir}
\abstract{In this paper, we focus on the possible correlation between conjectures KSS bound and weak gravity conjecture (WGC).  The hydrodynamic values KSS bound and weak gravity conjecture  constraint the low-energy effective field theory. These conjectures identify UV complete theories. We give  four, six and eight order derivative corrections to corresponding action and employ the hyperscaling violating charged AdS$_{d+2}$ black brane solution. These corrections lead us to find   correlation  between conjectures KSS bound and weak gravity conjecture.  We see that, with increasing perturbation correction, this correlation is more likely to appear. We consider dynamical constant $z=1$, $d=5$ and obtain the range of hyperscaling violation exponent $d+z-2\leq\theta\leq d+z-1$ for the above mentioned black brane. Here, we  show that higher derivative  corrections reduce the ratio of $\frac{M}{Q}$ to extremal black holes. Likewise, we also obtain the universal relaxation bound $\tau\geq \frac{1}{\pi T}$ and KSS bound $\frac{\eta}{s}\geq \frac{1}{4\pi}$ for our model. The results indicate that there is a possibility of a relationship between the two conjectures. Our studies also show the consistency of the WGC and the KSS bound conjectures for all corrections (except curvature-cubed, $\beta_2$) in the extremal and near-extremal condition.

 }

\begin{document}
\maketitle
\flushbottom

\section{Introduction}
\label{sec:intro}
As we know string theory proves that there is some extra dimension and the  physical properties of (3+1)-dimension phenomena depend on the choice of a small compact space. The existence of allowed small space give us many effective field theories in (3+1) dimension with consistence and inconsistence quantum gravity .  The effective field theory (EFT) that is consistent with quantum gravity is called the landscape. However, many effective theories are inconsistent and do not derive from string theory, they belong to the swampland. Such theory is an EFT but is not consistent with quantum gravity. How can we determine the difference between landscape and swampland and than say that such a theory is perfectly compatible with quantum gravity? There are some fundamental solution of EFT or some universal test to distinguish between the two theory as landscape and swampland. One of the criteria for the distinguishes of two theories is weak gravity conjecture (WGC). Here it means that gravity is always the weakest force \cite{1,2,3,4,5,6,7,8,9,10}.
Here we first try to explain WGC conjecture with the present of charged particle and black hole. As we know, the electric force between the elementary charge states is stronger than the gravitational attraction between them\cite{11,12,13,14,15,16,17}
\begin{equation}\label{eq:1.1}
(F_e=\frac{q^2}{r^2})\geq (F_g=\frac{m^2}{r^2}) \longrightarrow  q\geq m
\end{equation}
On the other hand, in the black hole, WGC is established in extremal conditions $M=Q$ and $T=0$. When a black hole does Hawking-radiation, the particles will out from it with mass $m$ and charge $q$. This means $Q-q\leq M-m$; as a result, we have $M=Q\longrightarrow q\geq m$. The mode of $m=q$ only occurs in the case of supersymmetry (the PBS state) and global symmetry. On the other hand, there are no such symmetries in the QG theory, so quantum corrections must remove the parameters of the black hole from their classical extremal values. In that case, even to extremal limit the corresponding corrections allow black holes to disappear. There are two possible predictions for quantum correction of black holes, $\frac{M}{Q}\leq 1$ and $\frac{M}{Q}\geq 1$. Calculations for various corrections show that the condition of $\frac{M}{Q}\leq 1$ is satisfied \cite{18,19,20,21,22,23,24,25,97}.

Here, to investigate the corresponding effective field theory and find the distinguishing of landscape and swampland we give some higher order derivative correction to the corresponding action. We know that Einstein's theory of gravity works well in the field of weak gravity, but when the scales become very small, the curvature becomes significant. At such scales, the Euclidean topological structure is improbable. The gravitational quantum fluctuations near the Planck length are so severe that they may impose a dynamical variable on the topological structure of the universe. The power of quantum effects on gravitational interactions in this range can be understood. As a result, we need to use generalized Lagrangian instead of Einstein-Hilbert Lagrangian. Here we note that higher-order terms are unconstrained terms and also independent of the original theory\cite{25,26,27,28,29,30,31,32,33,34,35,36,37,38,39,40,41}.

In this paper, we  use two special correction terms. Our first correction terms are Gauss-Bonnet, with two corrections related to charge ${\alpha}_{1}+{\alpha}_{3}$,
\begin{equation}\label{eq:1.2}
{\alpha}_{1}{R}^{2}+{\alpha}_{2}({R}^{2}_{abcd}+4{R}^{2}_{ab}+{R}^{2})+{\alpha}_{3}{R}^{2}_{ab}.
\end{equation}
and the second terms is curvature-cubed. These two corrections do not change under field invariant and do not break the supersymmetry. They are related to non-supersymmetry CFTs,
\begin{equation}\label{eq:1.3}
{\beta}_{1}({{R}^{ab}}_{mn}{{R}^{mn}}_{pc}{{R}^{pc}}_{ab})+{\beta}_{2}({{R}^{ab}}_{cd}{{R}^{cq}}_{pq}{{R}^{dp}}_{qb}).
\end{equation}
The last terms here is an eighth-order derivative that describes the expansion of general relativity with the string theory in sub-scale energies,
\begin{equation}\label{eq:1.4}
{\gamma}_{1}{({R}_{bcdm}{R}^{bcdm})}^{2}+{\gamma}_{2}{({R}_{bcdm}{\bar{R}}^{bcdm})}^{2}+{\gamma}_{3}({R}_{bcdm}{\bar{R}}^{bcdm}{R}_{bcdm}{R}^{bcdm}).
\end{equation}
For simplicity, we skipped the Maxwell field corrections in the corresponding model and postponed it to work later. All the above corrections at Planck scale are essential for the describing of full QG . However, a small area can be considered where these reforms prevail. By calculating the various constraints, we take a step towards the quantum mechanical regime. In that case we take advantage from  Abbot-Deser-Tekin method (ATD)and obtain the gravitational mass of these corrections. This method is fully described in subsection  \ref{subsec:3.1}.

Another case for distinguishing landscape from swampland is to investigate some hydrodynamic values in
the asymptotically AdS space-time. These values are presented as a conjecture for the holographic
field theory of dual. As we know, AdS/CFT conjecture is coming from holography and black hole objects play
 an important role in such conjecture. On the other hand, due to Hawking-radiation black hole objects follow
 statistical mechanics, so they have temperature and entropy. Quantum mechanics can be said to be a diffusion
  and attenuate theory at the fundamental level. It can be said that the asymptotically AdS space-time with event
  horizon is interpreted as thermal states in the field theory of dual, this means that the slight perturbations
   of a black hole or black brane correspond to minor deviations from the thermodynamic equilibrium in the field
    theory of dual. Our corrections lead to perturbation in the behavior of the black hole and black brane.
    Also here we say that the disturbance causes a similar behavior to hydrodynamics
    \cite{42,43,44,45,46,47,48,49,50,51,52,53}. As we know, hydrodynamics is an effective
    theory that describes the system's dynamics at large distances and time scale. According
    to the general second law of thermodynamics (GSL), the universal relaxation bound $\tau$
    is related to the mean free path $\ell_{mtp}$, which is given by,

\begin{equation}\label{eq:1.5}
\ell_{mtp}\sim \tau \sim \frac{1}{T},
\end{equation}
$\ell_{mtp}$ limits the description of hydrodynamics. These holographic arguments help to
 link quantum field theory to gravity. As a result, GSL is a powerful law that connects quantum theory,
 gravity and thermodynamics with respect to each other. Another factor that characterizes the universal
 relaxation properties of perturbation fluids is the lower limit of $\frac{\eta}{s}\geq\frac{1}{4\pi}$ or
 KSS bound (Kovtun-Starinets-Son)\cite{54,55,56,57,58,59,60,61,62,63,64,65,66}. Fluids respond
  to perturbations in two ways: the first leads to the sound state, which is caused by
longitudinal oscillations, and the second leads to the shear viscosity state, which arises from transverse
oscillations. The shear viscosity state $\frac{\eta}{s}$ indicates how
close a fluid is to perfection. This value is equal to $\frac{\hbar}{4\pi k_B}$ for a set of the quantum field theory
with strong interactions whose dual description involves black holes in AdS space-time . This boundary is saturated
for the theory of boundary fields in finite't Hooft coupling and the number of colors $N_c$. Such theories are
Einstein's gravity dual; these corrections show that it increases $\frac{\eta}{s}$.\cite{2}.
Another proof of conjecture is obtained from the explicit calculations of the $\alpha'$ correction
leading to type \textrm{IIB} string theory, which is compacted into 5-dimensions\cite{2}.
Also, the above corrections show that $\frac{\eta}{s}$ can be violated\cite{2}.
The first time that $\frac{\eta}{s}$ does not disappear is the 6-derivative correction.
From thermal field theory point of view , the calculation of shear viscosity through Kubo's formula,
the poles of the stress-energy tensor correlation function and the equation of state $D=\frac{\eta}{sT}$
have the same results. We also compute the diffusion equation for our model. We relate the fluid behavior
of the boundary theory to the fluid of the  bulk \cite{2}. It seems that our knowledge of the
relationship between hydrodynamics and the physics of black holes is still incomplete, and this article
improves our understanding of this issue\cite{67,68,69,70,71,72,73,74,75,76,77,78,79,80,81}.

Another perturbation that causes the species' hydrodynamic behavior in black holes is the entry of Dilaton into it.
Another application of the AdS/CFT is the study of coupled systems near critical points. At these points,
the system has scale symmetry and is described by a CFT theory. In many physical systems, critical points
are represented by a dynamic scale, under which the space-time scale varies,\cite{82,83,84,85,86,87}.
\begin{equation}\label{eq:1.6}
{ds}^{2}={r}^{\frac{-2\theta}{d}}(-{r}^{2z}{dt}^{2}+\frac{{dr}^{2}}{{r}^{2}}+{r}^{2}{d\vec{x}}^{2}),
\end{equation}
$d$, $z$ and $\theta$ are the dimension of space-time, the dynamic constant, and hyperscaling violations parameter,
respectively. We consider $d=5$, $z=1$ and $d-2+z\leq\theta\leq d-1+z$ in our background. This metric is not
inherently relativity, so that it can be considered for a toy model. In the geometries of non-relativistic
hyperscaling, the violation can also be considered an effective holographic description that lives in a finite
 branch $r$. The QFT states show its boundary in UV cutoff. For such theories, Einstein's gravity must be
 include the Abelian gauge field. In this article, we investigate Einstein-Hilbert-Maxwell-Dilaton's theory (EHMD).
  This is a reasonable extension of the scalar-tensor theory, which is based on the low-energy approximation of
  string theory and  its space based on the geometries of Lifshitz-like black brane\cite{88,89,90,91,92,93,94,95,96}.

All above mentioned information give us motivation to organize paper as follow. In section 2,
we describe the action and solution of the equations corresponding to of matter Lagrangian $\epsilon\epsilon$ .
In section 3, we first obtain the gravitational mass from the ADT method for the above mentioned corrections,
then we calculate the mass-to-charge ratio for all terms of  action. In section 4,
 we calculate the hydrodynamic values, including the ratio of shear viscosity to entropy density, etc.
 In section 5, we specify the common constraint of  WGC and  KSS bound, for all corrections.
 In the last section 6, we describe the results of our work.
\section{The Einstein-Hilbert-Maxwell-Dilaton  with corrections}
\label{sec:2}
In the beginning, to make exciting action for our model we use higher-order curvature corrections .
Using various references such as \cite{95,96}, we tried to get solution of the corresponding corrected action.

In that case, we show that the solution of action correspond to  hyperscaling violation charged
$AdS_{d+2}$ black
brane. On the other hand, such action can identified by the  constraints  of  commonalities of the
two theories as WGC and  KSS bound. So,  the above mentioned action with corresponding constraints lead us to obtain
two  parameters  of hyperscaling violation metric as $\theta$ and dynamical scaling $z$. Here, we face by
two part in action  as geometric with higher derivatives  ${\ell}_{G}$ and matter part of
Lagrangian as ${\ell}_{m}$. So, the corresponding action in $(d+2)$ dimension is given by,
\begin{equation}\label{eq:2.1}
\begin{split}
S=&\frac{1}{16\pi {G}_{d+2}}{\int}{d}^{d+2}x \sqrt{-g}(R-2\Lambda+{\ell}_{G}+{\ell}_{m}),\\
&{\ell}_{G}={\alpha}_{1}{R}^{2}+{\alpha}_{2}({R}^{2}_{abcd}+4{R}^{2}_{ab}+{R}^{2})+{\alpha}_{3}{R}^{2}_{ab}+{\beta}_{1}({{R}^{ab}}_{mn}{{R}^{mn}}_{pc}{{R}^{pc}}_{ab})+{\beta}_{2}({{R}^{ab}}_{cd}\\
&{{R}^{cq}}_{pq}{{R}^{dp}}_{qb})+{\gamma}_{1}{({R}_{bcdm}{R}^{bcdm})}^{2}+{\gamma}_{2}{({R}_{bcdm}{\bar{R}}^{bcdm})}^{2}+{\gamma}_{3}({R}_{bcdm}{\bar{R}}^{bcdm}{R}_{bcdm}{R}^{bcdm}),\\
&{\ell}_{m}=-\frac{1}{4}\sum_{i=1}^2{e}^{{\lambda}_{i}\phi}{F}^{2}_{i}-\frac{1}{2}{(\partial\phi)}^{2}+{\upsilon}_{0}{e}^{\upsilon\phi}\,,\qquad{\bar{R}}^{bcdm}={{\varepsilon}^{bc}}_{an}{R}^{andm}
\end{split}
\end{equation}
${\upsilon}_{0}$, ${\upsilon}$, ${\lambda}_{1,2}$ and $\Lambda$ are free parameters and cosmological
 constant respectively. In order to have quantum gravity in the Planck scale we essential to consider
 higher derivative correction terms.  We note here the couplings${\alpha}_{1,3}$, ${\alpha}_{2}$, ${\beta}_{1,2}$
 and ${\gamma}_{1,2,3}$ are charge, the Gauss-Bonnet, the curvature-cubed and curvature-quartile respectively.
 In the  matter part of  lagrangian, we need  to find gravitational theories in the framework of the $AdS/CFT$
 that describe the gravity of critical points (Like Lifshitz fixed points). Therefore, in this section,
 we consider the EHMD model. This model consists of a scalar field $\phi$ and two gauge fields $U(1)$,
 one for charge and the other along with scalar field generated an anisotropic scaling. We also use a
 typical exponential potential of string theory for the dilaton field. This potential is crucial to
 achieving a solution for this action and can be found near the horizons of different branes.
 We employ the equation of motion corresponding to matter Lagrangian ${\ell}_{M}$ \cite{95}.
\begin{equation}\label{eq:2.2}
\begin{split}
&\tau_{matter}=\frac{1}{2}\sum_{i=1}^2e^{\lambda_i\phi}({F^c_i}_{a}{F_i}_{cb}-{F}^{2}_{i}\frac{g_{ab}}{2d})+\frac{1}{2}{\partial_{a}\phi}{\partial_{b}\phi}-\upsilon_0e^{\upsilon\phi}\frac{g_{ab}}{d},\\
&\nabla^2\phi=-\upsilon_0e^{\upsilon\phi}\upsilon+\frac{1}{4}\sum_{i=1}^2\lambda_ie^{\lambda_i\phi}F^2_i\,,\qquad{\nabla}_{a}(\sqrt{-g}{e}^{{\lambda}_{i}\phi}{F}^{ab}_{i})=0,
\end{split}
\end{equation}
The solution of matter sector without the higher derivatives terms is given by,
\begin{equation}\label{eq:2.3}
\begin{split}
&{ds}^2=\frac{-r^{z}(r^{d+z}-Mr^{\theta}+Q^2r^{2(\theta+1)})}{r^{d+\frac{2\theta}{d}}}{dt}^2+\frac{r^{z+d-2-\frac{2\theta}{d}}}{(r^{z+d}-Mr^{\theta}+Q^2r^{2(\theta+1)})}{dr}^2+r^2{d\vec{x}}^2,\\
&F_2=Q r^{(\theta+1-d-z)} e^{\sqrt{\frac{1-z+\frac{\theta}{d}}{2(\theta-d)}}\varphi_{0}}\sqrt{2(\theta-d)(\theta+2-d-z)},\\
&e^{\phi}=e^{{\varphi}_0}r^{\sqrt{2(\theta-d)(1-z+\frac{\theta}{d})}}\,,\qquad\lambda_2=\sqrt{\frac{2(1-z+\frac{\theta}{d})}{\theta-d}},
\end{split}
\end{equation}
  where $K=0$ is flat space-time and $l=1$ is AdS radius. Here, $M$ and $Q$ are the mass and charge of the black hole,
  respectively. In the next section to include the terms of higher-derivatives ${\ell}_{G}$ in the solution,
  first one can  obtained the gravitational mass by the ADT method.

\section{The weak gravity conjecture}
\label{sec:3}
In this section, we are going  to calculate the extremality  condition of WGC. First, we use ADT method and
 obtain energy ${\ell}_{G}$,  then we couple  corresponding theory to the gauge and dilaton fields, and finally
 we calculate the ratio of $M/Q.$
\subsection{The gravitational mass with the ADT method}
\label{subsec:3.1}
Now we are to going briefly describe the ADT method. Our way is similar to the Landau-Lifshitz method, which is
asymptotically flat space-time. In this method, we obtain the equations of motion corresponding to the action.
 Here, the  gravitational theory is given by,
\begin{equation}\label{eq:3.1}
{\Psi}_{ab}(g,R,{R}^{2},{R}^{3},{R}^{4},...)=\omega{\tau}_{ab},
 \end{equation}
where $\omega$ and $\tau$ are gravitational coupling and stress-energy tensor of matter.
For linearization, we decompose the metric into ${g}_{ab}={\bar{g}}_{ab}+{h}_{ab}$. Metric
$\bar{g}_{ab}$ is to the equations (\ref{eq:2.1}) with ${\tau}_{ab}=0$, in this mode we have $E=0$ in this mode.
 Also here, the deviation ${h}_{ab}$ is for the linear state that relates to our gravitational terms and it
 disappears infinitely. The effective stress-tensor can be obtained by linearization,
\begin{equation}\label{eq:3.2}
{\Theta(\bar{g})}_{abcd}{h}^{cd}={\psi}_{ab}=\omega{T}_{ab},
 \end{equation}
where $\Theta(\bar{g})$ is the hermitian operator that depends on the background metric.
The equation (\ref{eq:2.2}) is obtained by the generalized Bianchi identity ${\bar{\nabla}}^{a}{\psi}_{ab}=0$,
the covariant conservation of the stress tensor ${\bar{\nabla}}^{a}{T}_{ab}=0$ and by placing ${g}_{ab}$.
Then, to calculate a conserved charge, we use the tensor ${T}_{ab}$. In this method, the gravitational mass
 (corresponding energy) can be obtained  by the  following integral,
\begin{equation}\label{eq:3.3}
E={\int}_{\varsigma}{d}^{d-1}x\sqrt{\bar{g}\varsigma} {n}_{a}{T}^{ab}{\bar{\xi}}_{b},
 \end{equation}
According to the above equation,  using time-like Killing vectors ${\bar{\xi}}_{b}$ and a constant-time
hypersurface ${\varsigma}$ with normal unit vector ${n}_{a}$, one can  write energy as  integral of bulk.
 We can now express the conserved current ${T}^{ab}{\bar{\xi}}_{b}$ as a complete derivative of potential
 ${\Gamma}^{ab}$, it means that ${T}^{ab}{\bar{\xi}}_{b}={\bar{\nabla}}_{a}{\Gamma}^{ab}$.
 The upper bulk integral can be written as a boundary integral, and ${r}_{b}$ is the boundary unit.
\begin{equation}\label{eq:3.4}
\textbf{E}={\int}_{\partial\varsigma}{d}^{d-2}x\sqrt{\bar{g}\partial\varsigma} {n}_{a}{r}_{b}{\Gamma}^{ab},
 \end{equation}
Now, we are going to calculate the corresponding energy (gravitational mass). As we know such energy
 is calculated for the six-derivative and the four-derivative theories by \cite{96}. Now, we take advantage from
 \cite{96} and obtain the above mentioned energy for the eight-derivative theory.
 The equation motion correspond to action (\ref{eq:2.1}) is obtained by the following expression,
\begin{equation}\label{eq:3.5}
-\frac{16\pi{G}_{d+2}}{\sqrt{-g}}\frac{\delta{S}_{m}}{\delta{g}^{ab}}={R}_{ab}-\frac{1}{2}{g}_{ab}R+\Lambda{g}_{ab}+{\psi}^{(i)}_{ab}=16\pi{G}_{d+2}{\tau}_{ab},
 \end{equation}
where ${\psi}^{(i)}_{ab}$ corresponds to the  equation of motion  for the i=4, 6, and 8-derivative corrections,
 which are given by,
\begin{equation}\label{eq:3.6}
\begin{split}
\psi^4_{ab}=&2\alpha_1R(R_{ab}-\frac{1}{4}Rg_{ab})+(2\alpha_1+\alpha_3)(-\nabla_b\nabla_a+g_{ab}g^{ab}\nabla_b\nabla_a)R+\alpha_3g^{ab}\nabla_b\nabla_a(R_{ab}\\
&-\frac{1}{2}Rg_{ab})+2\alpha_2[RR_{ab}-2R_{adbc}R^{dc}+R_{adcp}{R_b}^{dcp}-2R_{ad}{R^d}_b-\frac{1}{4}(R^2_{pfcd}-4R^2_{cd}\\
&+R^2)g_{ab}]+2\alpha_3(R_{adbc}-\frac{1}{4}R_{dc}g_{ab})R^{dc},
\end{split}
\end{equation}
\begin{equation}\label{eq:3.7}
\begin{split}
\psi^6_{ab}=&\frac{1}{2}[-6\beta_1\nabla_f\nabla_c({R_a}^{cpq}{{R_{pq}}^f}_b)+3\beta_2\nabla_f\nabla_c(R_{aqpb}R^{cqpf}-{R_a}^{qpf}{R^c}_{qpb})+\beta_1(3R_{dapc}\\
&R^{pcmn}{R^d}_{bmn}-\frac{1}{2}{R^{de}}_{pc}R^{pcmn}R_{mnde}g_{ab})+\beta_2(3{R^n}_{acd}{R^{cq}}_{pn}{{{R^d}_q}^p}_b-\frac{1}{2}{R^{ne}}_{cd}\\
&{R^{cq}}_{pn}{{{R^d}_q}^p}_eg_{ab})]+\frac{1}{2}[a\longleftrightarrow b],
\end{split}
\end{equation}
\begin{equation}\label{eq:3.8}
\begin{split}
\psi^8_{ab}=&\gamma_1(8R_{anbc}\nabla^n\nabla^cR_{bcdm}R^{bcdm}+\frac{g_{ab}}{2}(R_{bcdm}R^{bcdm})^2)+\gamma_2(8R_{apbn}\nabla^p\nabla^nR_{bcdm}{\bar{R}}^{bcdm}\\
&+(R_{bcdm}{\bar{R}}^{bcdm})^2\frac{g_{ab}}{2})+\gamma_3(4{\bar{R}}_{apbn}\nabla^p\nabla^nR_{bcdm}R^{bcdm}+4R_{apbn}\nabla^p\nabla^nR_{bcdm}{\bar{R}}^{bcdm}\\
&+\frac{g_{ab}}{2}R_{bcdm}{\bar{R}}^{bcdm}R_{bcdm}R^{bcdm}),
\end{split}
\end{equation}
Then, by placing the Riemann tensor, Ricci tensor and the scalar curvature in $d+2$-dimensions  in the above equations,
 we obtain   AdS solution with the effective cosmological constant \ref{eq:1A},
\begin{equation}\label{eq:3.9}
\begin{split}
\frac{32{(d+2)}^2(\gamma_3+\gamma_2+\gamma_1)}{d^4{(d+1)}^2}\Lambda^4_{eff}+\frac{4(d-4)(d\beta_2+4\beta_1)}{d^3{(d+1)}^2}\Lambda^3_{eff}\\
+\frac{2(d-2)(\alpha_3+\frac{d(d-1)}{(d+1)}\alpha_2+(d+2)\alpha_1)}{d^2}\Lambda^{2}_{eff}+{\Lambda}_{eff}=\Lambda
\end{split}
\end{equation}
Therefore, the perturbation solution of cosmological constant ${\Lambda}_{eff}$ is given by,
\begin{equation}\label{eq:3.10}
\begin{split}
{\Lambda}_{eff}=&\Lambda-\frac{2(2-d)(-\alpha_3-(2+d)\alpha_1)}{d^2}\Lambda^2-\frac{2(2-d)(1-d)\alpha_2}{d(2+d)}\Lambda^2+\frac{16(4-d)\beta_1}{d^3(1+d)^2}\Lambda^3\\
&+\frac{4(4-d)\beta_2}{d^2(1+d)^2}\Lambda^3-\frac{32(2+d)^2(\gamma_3+\gamma_2+\gamma_1)}{d^4(1+d)^2}\Lambda^4+\frac{8(-2+d)^2(\alpha_3+(2+d)\alpha_1)^2}{d^4}\\
&\Lambda^3+\frac{8(-1+d)^2(-2+d)^2\alpha^{2}_{2}}{d^2(1+d)^2}\Lambda^3+\frac{16(-1+d)(-2+d)^2(\alpha_3+(2+d)\alpha_1)\alpha_2}{d^3(1+d)}\Lambda^3\\
&+\frac{32(-4+d)^2(4\beta_1(\beta_2 d^2+4\beta_1)+d^2\beta^{2}_{2})}{d^6(1+d)^4}\Lambda^4+...,
\end{split}
\end{equation}
According to ${g}_{ab}={\bar{g}}_{ab}+{h}_{ab}$ and appendix \ref{app:1},  we linearize the equation of
motion and obtain the stress tensor ${T}_{ab}$,
\begin{equation}\label{eq:3.11}
\begin{split}
16G\pi{T}_{ab}=&\textbf{\bigg{[}}1+\frac{4((d+2)\alpha_1+\alpha_3)}{d}\Lambda_{eff}+\frac{4(1-d)(2-d)\alpha_2}{d(1+d)}\Lambda_{eff}-\frac{48(2d+1)\beta_1}{d^2(d+1)^2}\\
&\Lambda^{2}_{eff}+\frac{12(3d-2)\beta_2}{d(d+1)^2}\Lambda^{2}_{eff}+\frac{32(d+2)(\gamma_1-\gamma_2+\gamma_3)}{d^3(d+1)^2}\Lambda^{3}_{eff}\textbf{\bigg{]}}(R^L_{ab}-\frac{1}{2}{\bar{g}}_{ab}R_L\\
&-\frac{2\Lambda_{eff}}{d}h_{ab})+\textbf{\bigg{[}}2\alpha_1+\alpha_3+\frac{24\beta_1}{d(d+1)}\Lambda_{eff}+\frac{128(\gamma_1+\gamma_2+\gamma_3)}{d^2(d+1)^2}\Lambda^2_{eff}\textbf{\bigg{]}}({\bar{g}}_{ab}{\bar{g}}^{ab}\\
&{\bar{\nabla}}_a{\bar{\nabla}}_b-{\bar{\nabla}}_a{\bar{\nabla}}_b+\frac{2\Lambda_{eff}}{d}{\bar{g}}_{ab})R_L+\textbf{\bigg{[}}\alpha_3+\frac{6(8\beta_1-\beta_2)}{d(d+1)}\Lambda_{eff}\textbf{\bigg{]}}({\bar{g}}^{ab}{\bar{\nabla}}_a{\bar{\nabla}}_b(R^{L}_{ab}\\
&-\frac{1}{2}{\bar{g}}_{ab}R_L-\frac{2\Lambda_{eff}}{d}h_{ab})-\frac{2\Lambda_{eff}}{d}{\bar{g}}_{ab}R_L),
\end{split}
\end{equation}
${R}^{L}_{ab}$ and ${R}_{L}$ are the linearized  Ricci tensor and scalar curvature. As we have stated,
${\bar{\nabla}}^{a}{T}_{ab}=0$; therefore,  ${T}^{ab}{\bar{\xi}}_{b}={\bar{\nabla}}_{a}{\Gamma}^{ab}$,
we can obtain potential ${\Gamma}_{ab}$. Hence, the second and third terms of equation (\ref{eq:3.11})
will not affect on the  energy. And also, in our black brane model (\ref{eq:2.1}) will fall off at infinity.
It means that,
\begin{equation}\label{eq:3.12}
\begin{split}
&{\bar{\nabla}}^{a}({\bar{g}}_{ab}{\bar{g}}^{ab}{\bar{\nabla}}_{a}{\bar{\nabla}}_{b}-{\bar{\nabla}}_{a}{\bar{\nabla}}_{b}+\frac{2{\Lambda}_{eff}}{d}{\bar{g}}_{ab}){R}_{L}=0\\
&{\bar{\nabla}}^{a}({\bar{g}}^{ab}{\bar{\nabla}}_{a}{\bar{\nabla}}_{b}({R}^{L}_{ab}-\frac{1}{2}{\bar{g}}_{ab}{R}_{L}-\frac{2{\Lambda}_{eff}}{d}{h}_{ab})-\frac{2{\Lambda}_{eff}}{d}{\bar{g}}_{ab}{R}_{L})=0,
\end{split}
\end{equation}
So for the first term of equation (\ref{eq:3.11}),one can obtain the potential ${\Gamma}_{ab}$ as,
\begin{equation}\label{eq:3.13}
\begin{split}
16G\pi\Gamma^{ab}=&\textbf{\bigg{[}}1+\frac{4((d+2)\alpha_1+\alpha_3)}{d}\Lambda_{eff}+\frac{4(d-1)(d-2)\alpha_2}{d(d+1)}\Lambda_{eff}-\frac{48(2d+1)\beta_1}{d^2(d+1)^2}\Lambda^{2}_{eff}\\
+&\frac{12(3d-2)\beta_2}{d(d+1)^2}\Lambda^{2}_{eff}+\frac{32(d+2)(\gamma_1-\gamma_2+\gamma_3)}{d^3(d+1)^2}\Lambda^{3}_{eff}\textbf{\bigg{]}}({\bar{\xi}_p}\bar{\nabla^a}h^{bp}-{\bar{\xi}_p}\bar{\nabla^b}h^{ap}\\
-&{\bar{\xi}^b}\bar{{\nabla}^a}h+{\bar{\xi}^b}\bar{\nabla_p}h^{ap}+{\bar{\xi}^a}\bar{{\nabla}^b}h-{\bar{\xi}^a}\bar{{\nabla}_p}h^{bp}+h^{ap}\bar{{\nabla}^b}\bar{{\xi}_p}-h^{bp}\bar{{\nabla}^a}\bar{\xi_p}+h\bar{\nabla^a}\bar{\xi^b}),
\end{split}
\end{equation}
We note here that the energy is located in the couplings. Finally, using equations (\ref{eq:3.4}),
(\ref{eq:3.10}) and (\ref{eq:3.13}), we can obtain the  energy of perturbation case, which is given by.
\begin{equation}\label{eq:3.14}
\begin{split}
E=&\frac{md}{32G\pi}(1+\frac{4((d+2)\alpha_1+\alpha_3)}{d}\Lambda_{eff}+\frac{4(1-d)(2-d)\alpha_2}{d(1+d)}\Lambda_{eff}-\frac{48(2d+1)\beta_1}{d^2(d+1)^2}\Lambda^{2}_{eff}\\
+&\frac{12(3d-2)\beta_2}{d(d+1)^2}\Lambda^{2}_{eff}+\frac{32(d+2)(\gamma_1-\gamma_2+\gamma_3)}{d^3(d+1)^2}\Lambda^{3}_{eff}),
\end{split}
\end{equation}
where ${E}_{0}=\frac{md}{32G\pi}$ is  the case of non-correction of action.  Due to the perturbation of
equation (\ref{eq:3.11})
, our energy is also corrected by the corresponding terms.
\subsection{The Ratio $M/Q$ }
\label{subsec:3.2}
To obtain a ratio of $M$ to $Q$, we consider the following general ansatz,
\begin{equation}\label{eq:3.15}
{ds}^2=g_{tt}{dt}^2+g_{rr}{dr}^2+g_{xx}{d\vec{x}}^2
\end{equation}
The solution of equation (\ref{eq:2.1}) without corrections is given by,
\begin{equation}\label{eq:3.16}
\begin{split}
g_{tt}=&r^{4+2z-\theta}(-1+Mr^{2(\theta-\frac{d}{2}-\frac{\theta}{d})}-Q^2r^{2(1-d-z+3\frac{\theta}{2}-\frac{\theta}{d})}),\\
g_{rr}=&{r}^{\frac{-2\theta}{d}}({r}^{6}-M{r}^{(6-d-z+{\theta})}+{Q}^{2}{r}^{2(4-d-z+{\theta})})^{-1},\\
g_{xx}=&r^{2(1-\frac{\theta}{d})},
\end{split}
\end{equation}
The corrections solution  is fully described in appendix \ref{app:2}. For the non-correction mode,
we take $r_h$ as the radius of the outer horizon and put it in $f(r)$ (\ref{eq:2.3}) to
get $M(r_h)=r^{1/2}(1+Q^2r_h)$. Then from $f(r_h,r)=0$, we
calculate ${Q_1}^2=\frac{-r^{d+z-\theta}+{r_h}^{d+z-\theta}}{r^{2-d-z+\theta}-{r_h}^{2-d-z+\theta}}$.
 Hawking's temperature can be obtained by general formula of black hole, which is given by,
\begin{equation}\label{eq:3.17}
T=-\frac{\partial_r g_{tt}}{4\pi \sqrt{-g_{tt}g_{rr}}}{\bigg |}_{r=r_h}
\end{equation}
In that case the corresponding temperature for non-correction mode will be as,
\begin{equation}\label{eq:3.18}
T_{(0)}=\frac{r^{4+z}}{2\pi}(2+z-\frac{2\theta}{d}+M \frac{r^{\theta-d-z}}{2}(d-z-4+\frac{2\theta}{d}-\theta)-Q^2r^{2(\theta+1-d-z)}(d-3+\frac{\theta}{d}-\theta))
\end{equation}
By placing $M(r_h)$ obtained from (\ref{eq:2.3}) and examining the extremal condition $T=0$,
\begin{equation}\label{eq:3.19}
{Q_2}^2=\frac{-r^{d+z-\theta}(\theta-2d-dz)(d^2+\theta-d(3+\theta))+{r_h}^{d-\theta+z}(d^2+2\theta-d(4+\theta+z))}{2r^{2-d-z+\theta}(d^2+\theta-d(3+\theta))-{r_h}^{2-d+\theta-z}(d^2+2\theta-d(4+\theta+z))}
\end{equation}
In the extremal case, by setting ${Q_1}^2={Q_2}^2$, the $\theta$ range is obtained $d-2+z\leq\theta\leq d-1+z$.
This range is where the outer and inner horizons coincide. Now again we write the non-correction temperature
 for $d=5$, $z=1$ and on average of $\theta=4.5$, also we place $r=r_h$ and $M=r^{1/2}(1+Q^2r)$.
\begin{equation}\label{eq:3.20}
T_{0}(r,Q)=\frac{r^4}{8\pi}(Q^2+3r)
\end{equation}
The extremal condition is setting to $Q^2=-3r$. We also calculated the Hawking temperature of the
corrected solution, and we have given it in appendix \ref{app:3}.

On the other hand, we know that in the dual CFT, the temperature and time must be set to $T_{CFT}\equiv l_{eff}$
and $t\rightarrow t/l_{eff}$. Also, the energy density of the field theory is obtained by $M\equiv l_{eff}E$.
To compare the corrected solutions with the non-corrected ones, the temperature must not change at higher
order-correction. It means that $\frac{M}{Q}$ is written in terms of $T_{CFT}$ instead of $r$. To do this,
 we fix the new parameter $\tilde{r}$ in $T_{CFT}$.
\begin{equation}\label{eq:3.21}
T_{CFT}(\tilde{r},Q)=\frac{\tilde{r}^4}{8\pi}(Q^2+3\tilde{r})
\end{equation}
We get $r$ based on $\tilde{r}$ to remove $r$ in all terms in favor of $\tilde{r}$ by solving equation
$T_{CFT}\equiv l_{eff}T$. Then we make sure that the corrected and non-corrected solutions have the same
 temperature. The non-corrected mode is $r=\tilde{r}$, and for the rest of the higher order-derivative
  terms, it is calculated similarly in the extermal solution, $Q^2=-3\tilde{r}$.

Now we calculate ${(\frac{M}{Q})}_{0}$ for the non-correction state. Mass is obtained through the
 metric (\ref{eq:3.16}) and perturbation energy $E_0$ (\ref{eq:3.14}),
 $M_0=\frac{5 \tilde{r}^{1/2}}{32G\pi}(Q^2+\tilde{r})$. The charge is obtained for
 ${\Gamma}^{ab}=\sqrt{-g}{e}^{{\lambda}_{i}\phi}{F}^{ab}_{i}$ and (\ref{eq:2.3})
  by solving integral (\ref{eq:3.4}), $Q_0=\frac{i}{\sqrt{2}}Q \tilde{r}^{(\frac{3i}{\sqrt{10}}-\frac{1}{2})}
  \exp(\frac{9i\varphi_0}{\sqrt{10}})$. As a result,
\begin{equation}\label{eq:3.22}
{(\frac{M}{Q})}_{0}=\frac{5\sqrt{2} \exp(\frac{9}{\sqrt{-10}}\varphi_0)}{32 G\pi i}(Q^2+\tilde{r})\frac{\tilde{r}^{\frac{3}{\sqrt{-10}}+1}}{Q}
\end{equation}
Hence, all correction  terms are calculated by appendix \ref{app:2} with placing $(\tilde{r}=1)$.
\begin{equation}\label{eq:3.23}
\begin{split}
\frac{M}{Q}=&{(\frac{M}{Q})}_{0}(1-\alpha_1\big{[}\frac{42(291.6+459.7Q^2+177Q^4+8.4Q^6)}{291.01+566.4Q^2+295.2Q^4+19.8Q^6}\big{]}-\alpha_2\big{[}\frac{48(132+139Q^2+7Q^4)}{264+287Q^2+23Q^4}\big{]}\\
-&\alpha_3\big{[}\frac{192(1+1.5Q^2+0.6Q^4+0.05Q^6)}{15.6+32.1Q^2+19Q^4+2.5Q^6}\big{]}-\beta_1\big{[}\frac{5376(1+1.3Q^2+0.07Q^4-0.1Q^6)}{70.1+87.2Q^2-23.7Q^4-40.8Q^6}\big{]}\\
+&\beta_2\big{[}\frac{4160(1+1.5Q^2+0.5Q^4+0.02Q^6)}{32+61.7Q^2+31.3Q^4+1.6Q^6}\big{]}+\gamma_2\big{[}\frac{168(24.2+38.9Q^2+9.9Q^4-1.5Q^6)}{27.5+43.8Q^2+10.08Q^4-6.1Q^6}\big{]}\\
-&(\gamma_1+\gamma_3)\big{[}\frac{168(12.9+20.4Q^2+7.9Q^4+0.4Q^6)}{291.01+566.4Q^2+295.2Q^4+19.8Q^6}\big{]}+.)
\end{split}
\end{equation}
At extremality case as  $Q^2=-3$, one can write following expression,
\begin{equation}\label{eq:3.24}
\frac{M}{Q}=\frac{4.08}{16 G \pi}e^{-2.84 i \varphi_0}(1-16.3\alpha_1-27.3\alpha_2-6.6\alpha_3-20.06\beta_1+48.8\beta_2-68.4(\gamma_1+\gamma_3)+42.1\gamma_2+.)
\end{equation}
The  perturbation of  black holes in the extremal and near-extremal states is satisfied
 by the $\frac{M}{Q}\leq 1$ condition . Our results show that the establishment of this condition
  depends on the sign of the couplings. In section \ref{sec:5}, we examine both $\alpha_1>0$
  and $\alpha_1<0 $ modes for all couplings.

\section{The Hydrodynamics}
\label{sec:4}
As a mentioned before another approach of distinguishing a landscape from swampland is  KSS bound which is
coming from hydrodynamic side. For this reason in this section we describe hydrodynamics and its values.
\subsection{Limits of thermodynamic}
\label{subsec:4.1}
As we know, the second law of thermodynamic is the bridge between quantum theory, gravity and thermodynamics.
 According to it, the entropy of $S_{matter}+ S_{hole}$ never decreases in the interaction between matter and
  black hole. It predicts the quantity of entropy  generally will be universal. We note here, this corresponding
  law creates two quantum constraints.
\textbf{I. The minimum length scale:}\\
The first constraint coming from minimum length scale and there is a specific length range to the thermodynamic system.
 It is $\ell_{min}=\frac{1}{2\pi T}$ for a fluid with zero chemical potential. So,  our corrections lead us to
 have following expression,
\begin{equation}\label{eq:4.1}
\ell_{min}=(9\times10^{15}-342784.2\alpha_1-0.3\alpha_3-0.09\beta_1-159.08i\beta_2+1467.5i(\gamma_1+\gamma_3)+311229i\gamma_2)
\end{equation}
The temperature for term $\alpha_2$ is zero, which means that we have infinite length. Generally,
 one can say that here hydrodynamics is an effective theory for perturbation of Gauss-Bonnet  at all
 lengths. Other above values also indicate the effectiveness of the hydrodynamics theory for perturbations
  of hyperscaling violating charged AdS$_{d+2}$ black brane.

\paragraph{II. Universal relaxation bound:}
Another constraint coming from  $\tau\geq\frac{1}{\pi T}$ which  is the time limit of a perturbed thermodynamic
system to achieve universal relaxation. This constraint for our corrections,
\begin{equation}\label{eq:4.2}
\tau=(1.8\times10^{16}-685568.4\alpha_1-0.6\alpha_3-0.18\beta_1-318.1i\beta_2+2935.08i(\gamma_1+\gamma_3)+622458.1i\gamma_2+.)
\end{equation}
So, this above limits of length  give us the validity of the effective hydrodynamic description.
On the other hand, each fluid has two types of physical responses to different perturbations,
shear viscosity and sound. The most extended turbulence state has  wave-length
as $\lambda_{max}=2\pi \ell$ and wave-number as $\kappa_{min}=\frac{2\pi}{\lambda}=2\pi T$.
In the following  we have two parts, first we calculate the shear mode $(\frac{\eta}{s}$ for
the correction case, and then second part we obtain its diffusion relation.
\subsection{The KSS Bounds}
\label{subsec:4.2}
This section we calculate the ratio of shear viscosity to entropy density for hyperscaling violation
charged AdS$_{d+2}$ black brane with higher-derivative corrections. The shear viscosity is
 given by Kubo's formula with  two-point correlation function of the stress-energy tensor will be as,
\begin{equation}\label{eq:4.3}
\eta=\lim_{\omega \longrightarrow \infty} \int d^4\vec{x} \frac{e^{i\omega t}}{2\omega} \langle[T_{xy}(x),T_{xy}(0)]\rangle
\end{equation}
We express the two-point function as a retarded Green's function of $T_{xy}$,
\begin{equation}\label{eq:4.4}
G_{tx,tx}(\omega,q)=-i\int \Theta(t) d^4\vec{x} e^{i(\omega t-qz)} {\langle[T_{tx}(x),T_{tx}(0)]\rangle} \propto (-D q^2+i\omega)^{-1}
\end{equation}
It has a pole at $\omega=-i D q^2$. The shear diffusion constant $D$ is related to the
plasma entropy density of the gauge theory, which is given by,
\begin{equation}\label{eq:4.5}
D=\frac{\eta}{sT}
\end{equation}
In the thermal field theory, the calculation of shear viscosity via all three of the above equations has the
same results. According to the temperature calculation in subsection \ref{subsec:3.2}, we obtain the
shear viscosity from equation (\ref{eq:4.5}). We have given the temperature (\ref{eq:3.17})
for all correction terms in appendix \ref{app:3}. At present, with the corrected solutions \ref{app:2}, we can acquire the diffusion constant for all corrective terms,
\begin{equation}\label{eq:4.6}
D=\sqrt{\frac{g(r_h)}{g_{tt}(r_h)g_{rr}(r_h)}}\int_{r_h}^{\infty} dr \frac{-g_{tt}g_{rr}}{g_{xx}\sqrt{-g}}
\end{equation}
We can now examine the universality of the shear viscosity ratio to the entropy density $\frac{\eta}{s}=TD$
for our action  . In case of  non-corrective state of the hyperscaling violation charged AdS$_{d+2}$ black brane,
 this ratio is,
\begin{equation}\label{eq:4.7}
\frac{\eta}{s}=\frac{1}{4\pi}(-2.45Mr^{5/2}+2.9Q^2r^3+3.81r^4)
\end{equation}
For $r=r_h$, $d=5$ and $z=1$, we get an average of $\theta=4.5$.  Using equations (\ref{eq:4.6}), appendix \ref{app:2}
 and \ref{app:3} for extremal conditions, we  obtain the ratio $\frac{\eta}{s}$. In appendix \ref{app:4},
  we also provide  $\frac{\eta}{s}$ for all above mentioned corrections to the system .
\begin{equation}\label{eq:4.8}
\begin{split}
\frac{\eta}{s}=&\frac{1}{4\pi}(0.014+0.0001\alpha_1-0.0000008\alpha_3+0.000001\beta_1-6893.63\beta_2\\
+&0.002\gamma_1-0.00005\gamma_2+0.002\gamma_3+..)\simeq\frac{1}{4\pi}(-6893.63\beta_2)
\end{split}
\end{equation}
 The above results indicate the violation of condition KSS bound in different couplings except for $\beta_2$,
 for this reason we consider only $\beta_2$. We thoroughly compare corrections to the KSS bound and the
 WGC for all coupling marks in section \ref{sec:5}.
\subsection{The Dispersion Law}
\label{subsec:4.3}
Shear viscosity is the inherent ability of a turbulent fluid to relax toward equilibrium,
so there is a possibility of a relationship between KSS bound and a low thermodynamic limit
of the universal relaxation bound. We express this relationship with the help of the diffusion equation.
\begin{equation}\label{eq:4.9}
\omega=i\frac{\eta}{sT}\kappa^2+i\tau(\frac{\eta}{sT})^2\kappa^4+...\cong0
\end{equation}
According to equations (\ref{eq:4.1}), (\ref{eq:4.5}) and wave-number, one  can be obtain for
their corrections $\omega\cong0$. This value of diffusion determines the inherent ability of
a fluid to eliminate perturbation and approach thermal equilibrium.

\section{The connection WGC and KSS}
\label{sec:5}
We now examine the results of weak gravity conjecture and hydrodynamics for the corresponding
 corrections. Also here we show that under what conditions they have relationship to each other.
  To better perception the common constraint, we categorized the perturbations into three sections as follows:

\subsection{Four-Derivatives}
\label{sec:5.1}
We first consider the case where the perturbations for the six and eight order-derivatives are off,
i.e.$\beta_{1,2}=\gamma_{1,2,3}=0$. The terms of this part is  satisfied by the supersymmetry.
 Modes $\alpha_1$, $\alpha_2$, and $\alpha_3$ are created by the energy of effective action
 correspond to the heterotic string. According to (\ref{eq:3.24}) and (\ref{eq:4.8}),
 for the corresponding above conditions in the extremal condition, we have

\begin{equation}\label{eq:5.1}
\frac{M}{Q}=(\frac{M}{Q})_0(1-16.38\alpha_1-27.32\alpha_2-6.6\alpha_3)\,,\qquad\frac{\eta}{s}=\frac{1}{4\pi}
\end{equation}
These results show that, the common range between two conjectures as WGC and KSS bound
in the extremal conditions is equal to $\alpha_1=0.061$, $\alpha_2=0.036$ and $\alpha_3=0.151$.

\begin{figure}[tbp]
\centering
\includegraphics[width=.23\textwidth,origin=c,angle=0]{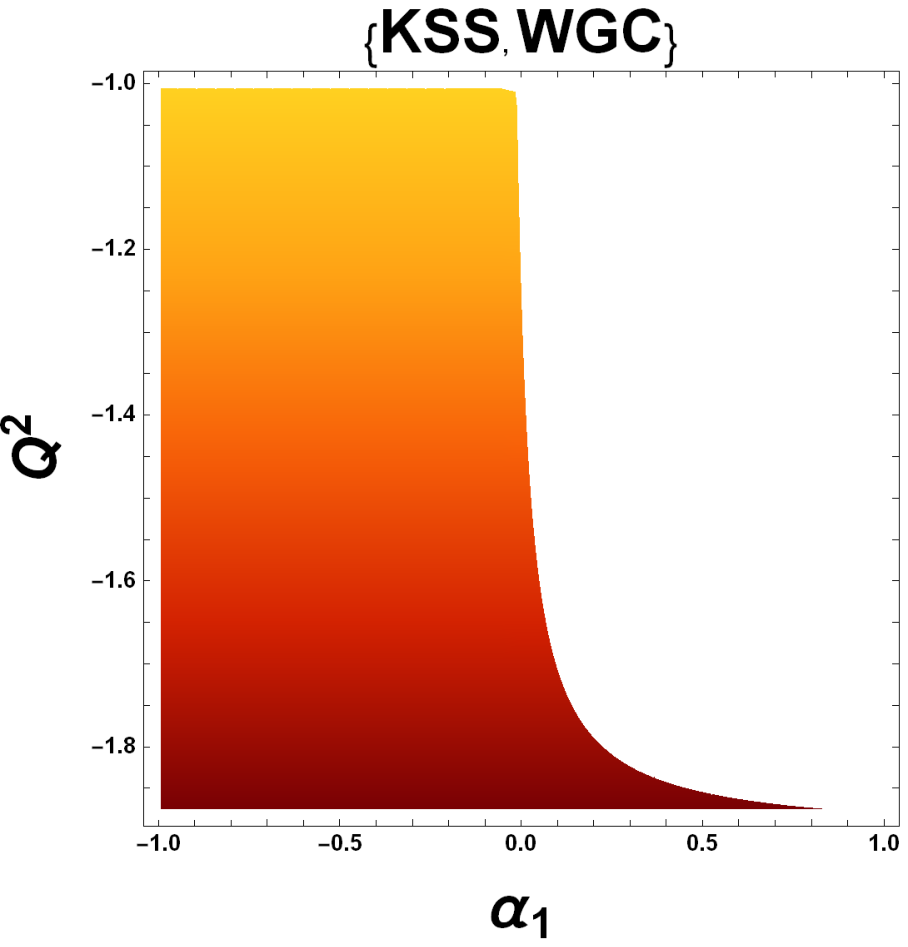}
 \hfil
 \includegraphics[width=.23\textwidth,origin=c,angle=0]{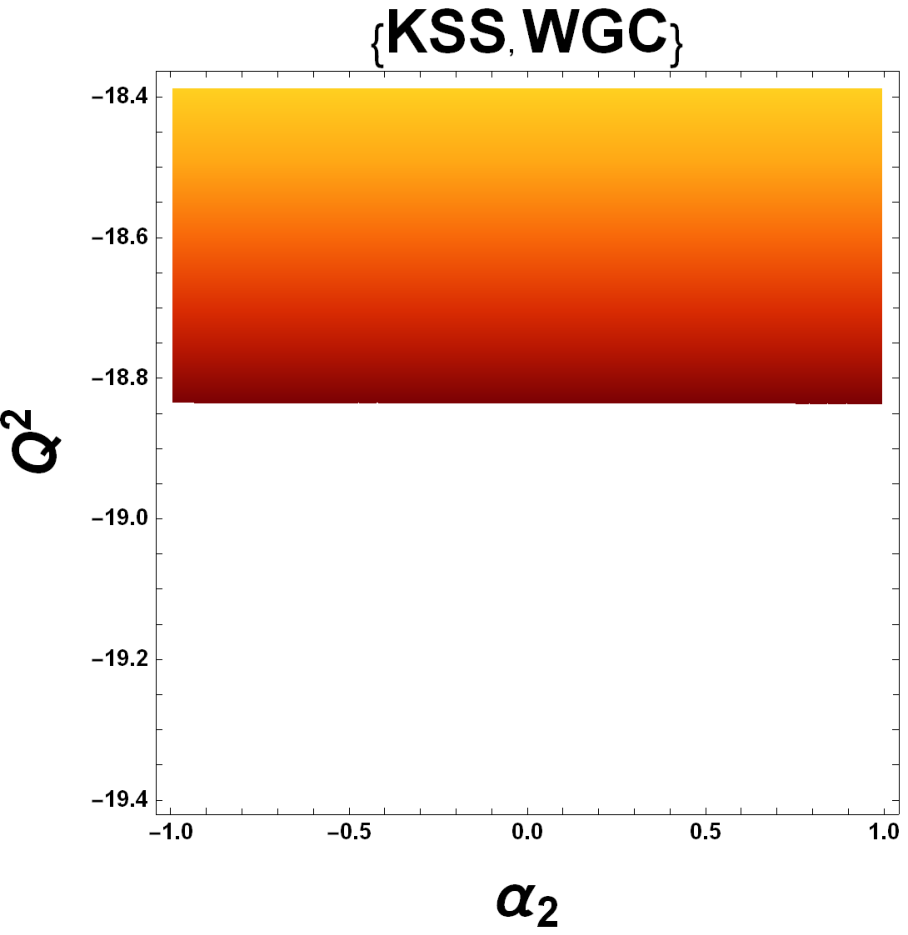}
\hfil
 \includegraphics[width=.23\textwidth,origin=c,angle=0]{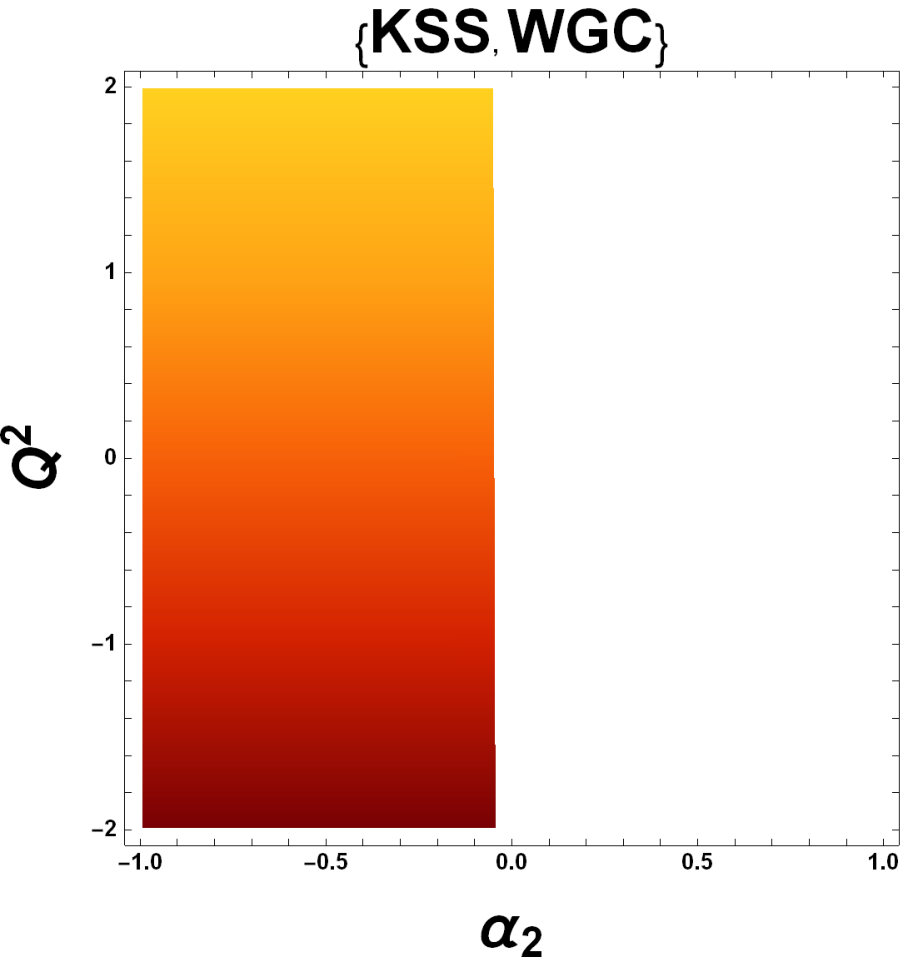}
 \hfil
 \includegraphics[width=.23\textwidth,origin=c,angle=0]{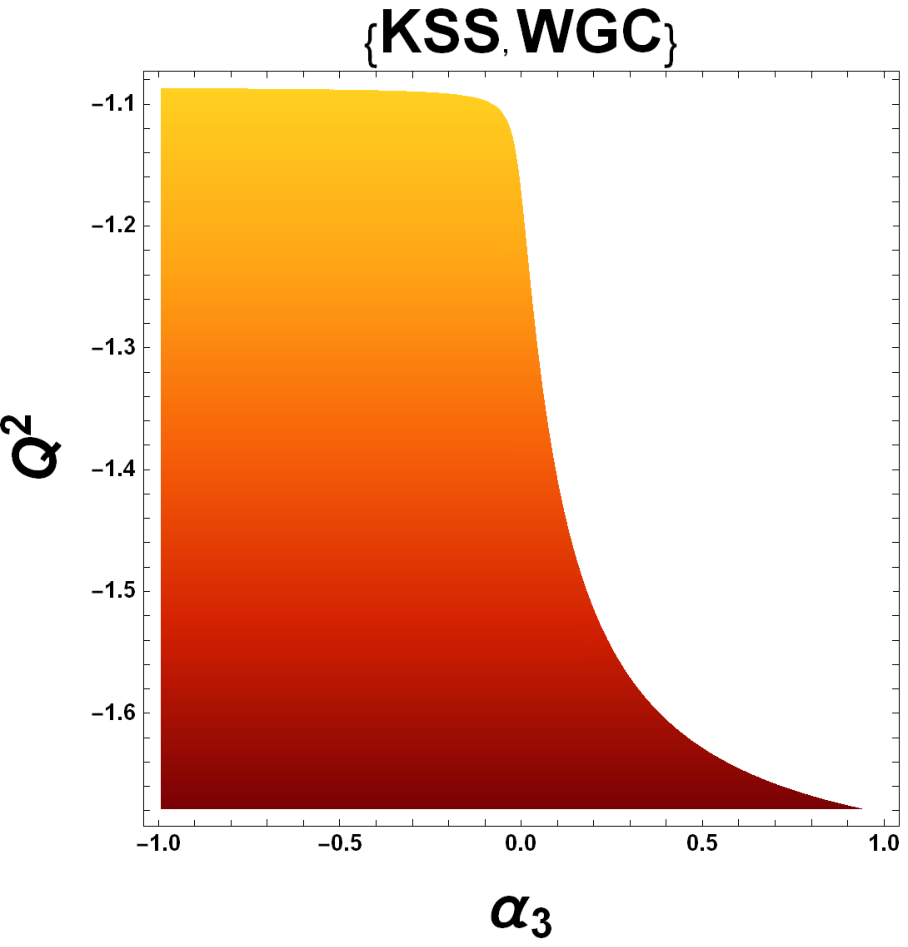}
\caption{\label{fig:1}Plot of $(\alpha_i-Q^2)$. The solar colors' region represent where both the KSS bound and the WGC are satisfied, for different couplings of 4-derivative perturbations.}
\end{figure}

\begin{table}[tbp]
 \centering
 \begin{tabular}{|c|c|c|}
 \hline
 $Coupling$     & $Charge-range$             & $Common-range$              \\
 \hline
 $\alpha_1>0$   &$-1.877\leq Q^2 \leq-1.003$ & $-1.877\leq Q^2 \leq-1.003$ \\
 $\alpha_1<0$   &$-1.934\leq Q^2 \leq-1.003$ &                             \\
 \hline
 $\alpha_2>0$   &$Q^2=-18.38$                & $-19.39\leq Q^2 \leq-18.38$ \\
 $\alpha_2<0$   &$Q^2=-19.39$                &                             \\
 \hline
 $\alpha_3>0$   &$ -1.68\leq Q^2 \leq-1.083$ & $-1.68\leq Q^2 \leq-1.085$  \\
 $\alpha_3<0$   &$-1.82\leq Q^2 \leq-1.085$  &                             \\
 \hline
 \end{tabular}
 \caption{\label{tab1}Charge range for 4-derivative couplings in near-extermal condition.}
 \end{table}
The range of KSS bound,  near-extremal condition (\ref{eq:3.23}) covers everywhere due to the low value of
 the ratio  shear viscosity to entropy density (\ref{eq:4.8}); therefore, the WGC space specifies the joint range.
 According to the sign of couplings, we have different charging ranges in table \ref{tab1}. The joint range of
 their charge gives us the common constraint the WGC and  KSS bound. In that case,  we have drawn  figure \ref{fig:1}.

We do not consider the values $-18.5\leq Q^2\leq-17.86$ for $\alpha_1$ and $-12.5\leq Q^2\leq-9.4$ for $\alpha_3$.
For $\alpha_2$, we also  saw the  results of $-2\leq Q^2 \leq2 $, were not in our answers. The third
figure \ref{fig:1} shows that even in this case, we have a joint range. Also, we considered the coupling
range to be $-1\leq \alpha_{1,2,3}\leq+1$. It should be noted that mode $\alpha_2$ is the Gauss-Bonnet
term, which we have shown for this mode; both conditions are satisfied within specific charging ranges.

\subsection{Six-Derivatives}
\label{subsec:5.2}
In the second part, we examine theories considered toy models for non-supersymmetry string theory compactifications.
Those theories may be dual to CFTs. These theories are the same corrections as the six order-derivatives
with $\beta_1$ and $\beta_2$. We consider the state $\alpha_{1,2,3}=\gamma_{1,2,3}=0$. First, we examine
 $\beta_1$, so we have equations (\ref{eq:3.24}) and (\ref{eq:4.8}) for the extremal conditions.
\begin{equation}\label{eq:5.2}
\frac{M}{Q}=(\frac{M}{Q})_0(1-20.06\beta_1)\,,\qquad\frac{\eta}{s}=\frac{1}{4\pi}
\end{equation}
In extremal condition,  two modes of the KSS bound and  WGC are satisfied by the point of  $\beta_1=0.049)$.

\begin{figure}[tbp]
\centering
\includegraphics[width=.32\textwidth,origin=c,angle=0]{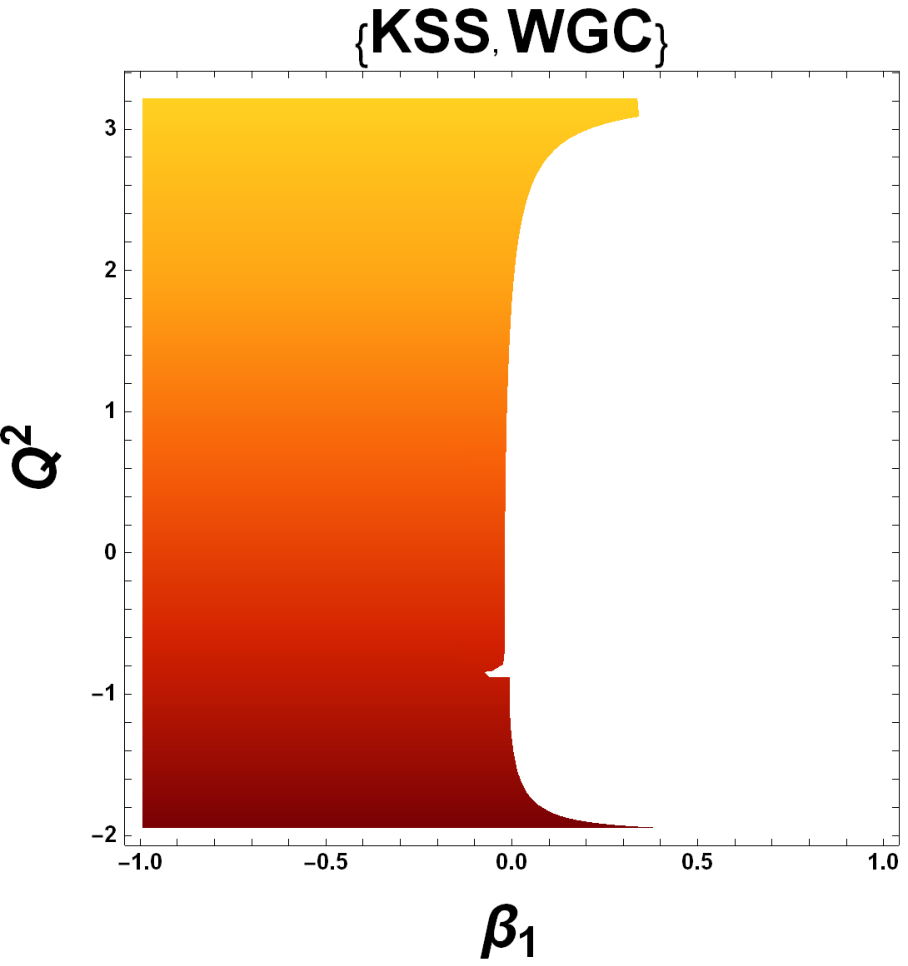}
 \hfil
 \includegraphics[width=.32\textwidth,origin=c,angle=0]{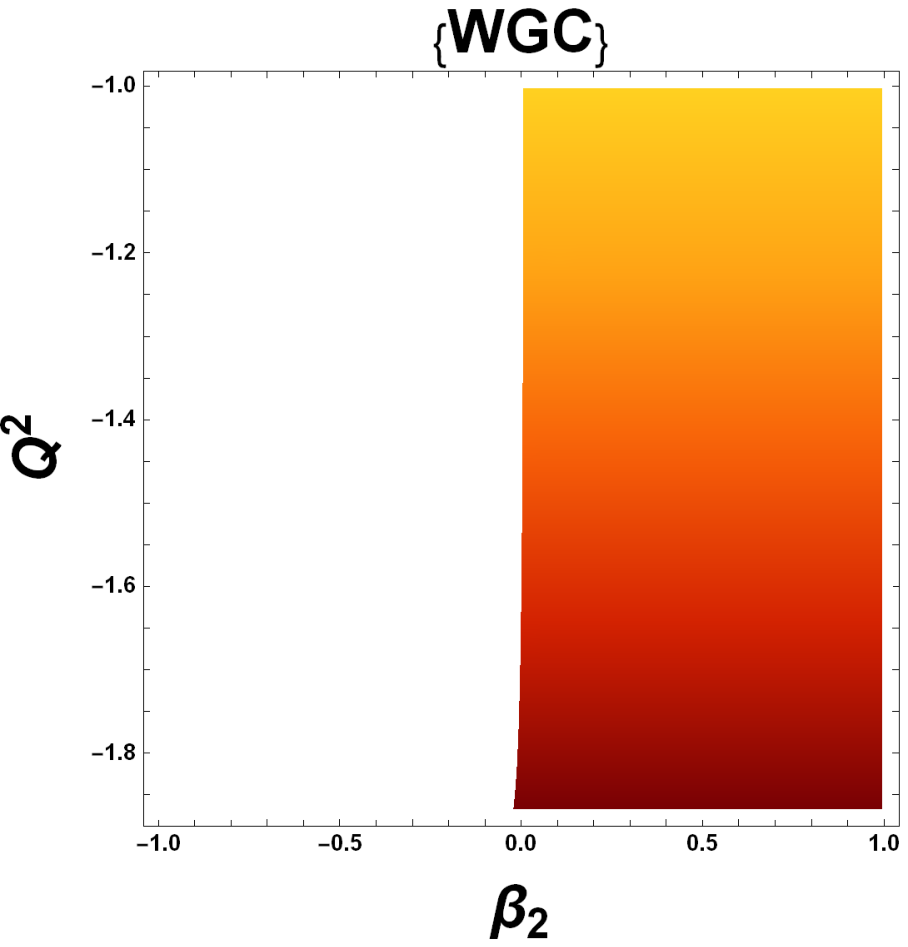}
\hfil
 \includegraphics[width=.32\textwidth,origin=c,angle=0]{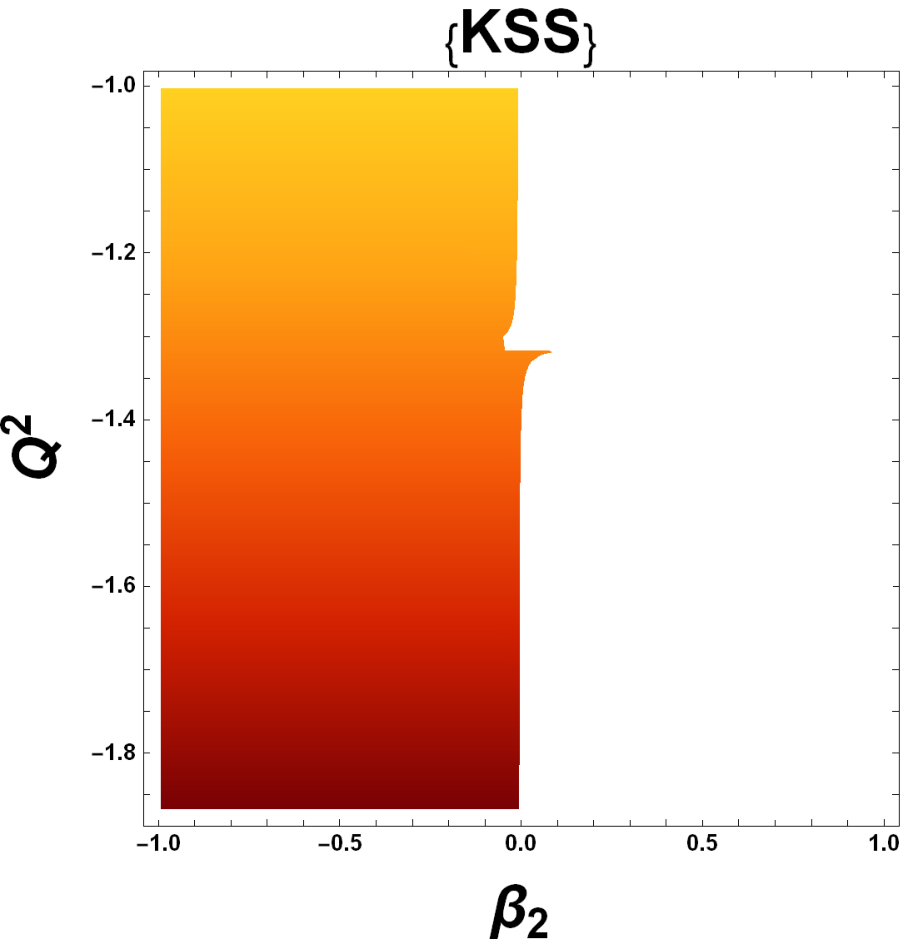}
\caption{\label{fig:2}Plot of $(\beta_i-Q^2)$. For $\beta_1$ the solar colors' region represent where both the KSS bound and the WGC are satisfied, but for $\beta_2$ show these two conjectures have no satisfy region.}
\end{figure}

\begin{table}[tbp]
  \centering
  \begin{tabular}{|c|c|c|c|}
  \hline
                 &$Coupling$    & $Charge-range$             & $Common-range$              \\
  \hline
     $WGC$       &$\beta_1>0$   &$-1.96\leq Q^2 \leq 3.176$  & $-1.96\leq Q^2 \leq 3.3$    \\
                 &$\beta_1<0$   &$-2.003\leq Q^2 \leq 3.3$   &                             \\
  \hline
      $WGC$      &$\beta_2>0$   &$-1.887\leq Q^2 \leq-0.999$ &  $-1.869\leq Q^2 \leq-1.00$ \\
                 &$\beta_2<0$   &$-1.869\leq Q^2 \leq-1.00$  &                             \\
  \hline
      $KSS$      &$\beta_2>0$   &$-2.88\leq Q^2 \leq2.88$    & $-2.88\leq Q^2 \leq2.88$    \\
                 &$\beta_2<0$   &$-2.88\leq Q^2 \leq2.88$    &                             \\
  \hline
  \end{tabular}
  \caption{\label{tab2}Charge range for 6-derivative couplings in near-extermal condition.}
  \end{table}
In near-extremal condition, the KSS bound is satisfied to all charges, so again, we only need to find the WGC range.
According to equation (\ref{eq:3.23}), we calculate the charge range for $\beta_1$ different sign and obtain
the joint range, see table \ref{tab2}. Then, to satisfy KSS bound and  WGC for the coupling $\beta_1$,
we draw figure \ref{fig:2}.
 According to equations (\ref{eq:3.24}) and (\ref{eq:4.8}), we see the affect of coupling  $\beta_2.$

\begin{equation}\label{eq:5.3}
\frac{M}{Q}=(\frac{M}{Q})_0(1-48.84\beta_2)\,,\qquad\frac{\eta}{s}=\frac{1}{4\pi}(-6893.63\beta_2)
\end{equation}
The above results show that the extremal condition of  KSS bound and  WGC are not satisfied together.
Here, checking the near-extremal condition can yield exciting results. We calculate the charge range separately
from equations (\ref{eq:3.23}) and (\ref{eq:5C}) and obtain their joint range equal to $-1.869\leq Q^2 \leq-1.00$,
 table \ref{tab2}. Then we draw them separately, figure \ref{fig:2}. According to this figure, there is no
 satisfying  between the KSS and the WGC in condition near-extremal. Note that we have checked mode
 $\theta=4.5$, $d=5$, $z=1$ at hyperscaling violating charged AdS$_{d+2}$ black brane. As we calculated in section
 \ref{sec:3}, $\theta$ has a range of $d-2+z\leq\theta\leq d-1+z$. It indicates that the possibility of
 compatibility of these conjectures with other values has not yet been eliminated.

\subsection{Eight-Derivatives}
\label{subsec:5.3}
The latest corrections reviewed are the eight order-derivatives. These corrections describe the development
of general relativity with string theory at sub-scale energies. In this case, too, we first consider the
extremal condition for all three terms of equations (\ref{eq:3.24}) and (\ref{eq:4.8})
with $\alpha_{1,2,3}=\beta_{1,2}=0$.

\begin{equation}\label{eq:5.4}
\frac{M}{Q}=(\frac{M}{Q})_0(1-68.43\gamma_1+42.18\gamma_2-68.43\gamma_3)\,,\qquad\frac{\eta}{s}=\frac{1}{4\pi}
\end{equation}
Conjectures of the KSS bound and the WGC are consistent in points of $\gamma_{1,3}=0.14$ and $\gamma_2=-0.023$.

\begin{figure}[tbp]
\centering
\includegraphics[width=.32\textwidth,origin=c,angle=0]{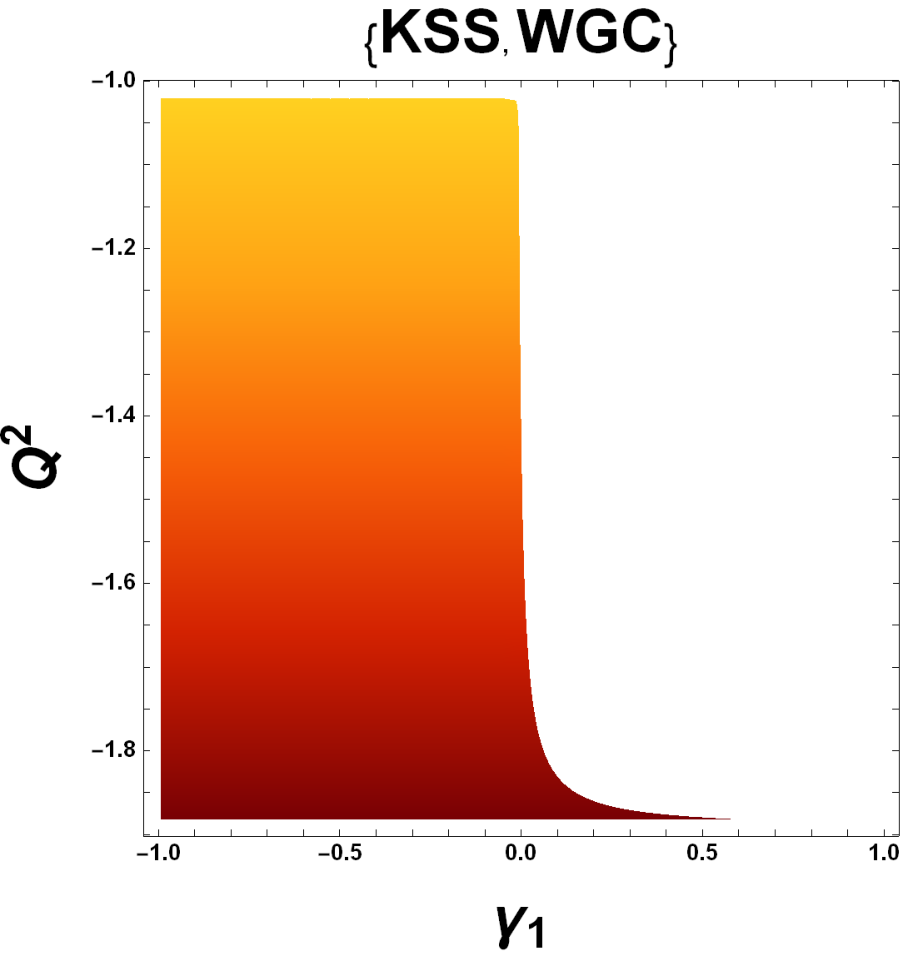}
 \hfil
 \includegraphics[width=.32\textwidth,origin=c,angle=0]{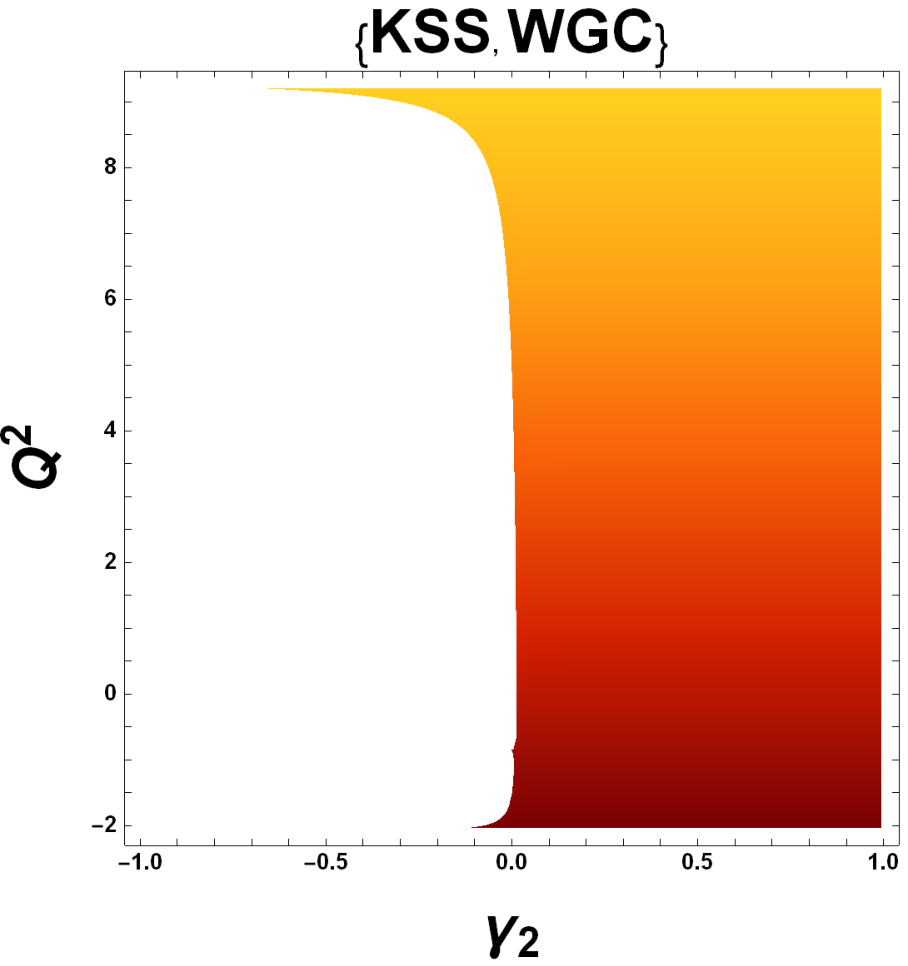}
\hfil
 \includegraphics[width=.32\textwidth,origin=c,angle=0]{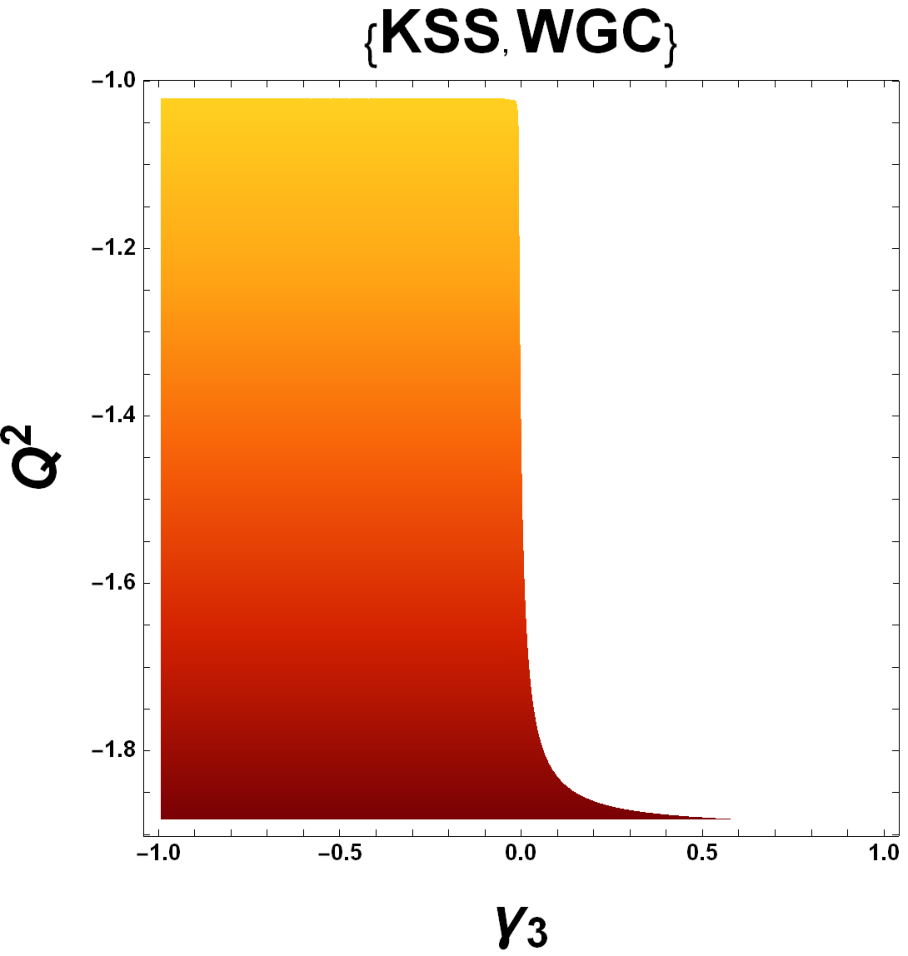}
\caption{\label{fig:3}Plot of $(\gamma_i-Q^2)$. The solar colors' region represent where the KSS bound and the WGC are satisfied.}
\end{figure}

\begin{table}[tbp]
  \centering
  \begin{tabular}{|c|c|c|}
  \hline
  $Coupling$    & $Charge-range$             & $Common-range$              \\
  \hline
  $\gamma_1>0$  &$-1.883\leq Q^2 \leq-1.017$ & $-1.883\leq Q^2 \leq-1.017$ \\
  $\gamma_1<0$  &$-1.897\leq Q^2 \leq-1.017$ &                             \\
  \hline
  $\gamma_2>0$  &$-2.083\leq Q^2 \leq9.480$  & $-2.064\leq Q^2 \leq9.232$  \\
  $\gamma_2<0$  &$-2.064\leq Q^2 \leq9.232$  &                             \\
  \hline
  $\gamma_3>0$  &$-1.883\leq Q^2 \leq-1.017$ & $-1.883\leq Q^2 \leq-1.017$ \\
  $\gamma_3<0$  &$-1.897\leq Q^2 \leq-1.017$ &                             \\
  \hline
  \end{tabular}
  \caption{\label{tab3}Charge range for 8-derivative couplings in near-extermal condition.}
  \end{table}
For the near-extremal condition, we calculate the charge range of all three couplings from equation
(\ref{eq:3.23}), Table \ref{tab3}. We do not consider $-15.73\leq Q^2\leq-15.56$
answers for $\gamma_1$ and $\gamma_3$. According to the common charge limits for
the near-extremal condition of two conjectures, the WGC and the KSS bound,
 we have drawn the consistent region for all couplings,  we see figure~\ref{fig:3}.

\section{Conclusions}\label{sec:6}
In this paper, we first calculated the gravitational mass for the higher derivative corrections
(toy-models (\ref{eq:2.1})) from the covariant ADT method in the AdS background. In this method,
gravitational mass (energy) can be obtained by linearizing the equations of motion without the
need for counter-terms. Also, the effective stress tensor (\ref{eq:3.11}) has maintained its state
in all terms and only its coefficients have changed. These results are equal to \cite{96}. Given that,
this effective stress tensor has nothing to do with the matter part of the action. It can be examined
whether this effective stress tensor can be a general measure for different modes in higher
corrections or not? The results of our study prove this for the first-order corrections in
the $4$, $6$, and $8$-derivatives.
We then obtained a ratio of $\frac{M}{Q}$ and show that higher derivative corrections reduce
the mass to charge ratio of the extremal black brane (WGC). We also examined the hydrodynamic
values of the black hole in section \ref{sec:4}. Likewise, we obtained the universal relaxation
bound $(\tau\geq \frac{1}{\pi T})$ by Hawking temperature, shear viscosity to entropy ratio
$(\frac{\eta}{s}\geq \frac{1}{4\pi})$ by shear diffusion equation and dispersion equation
$\omega$ of these hydrodynamic values. Our result shown that there is a possibility of
violating these values in higher corrections. It also seems that probability there is a
relationship between the universal relaxation bound and KSS bound. Such constraints helped us
to distinguish which theories can be UV completed. Finally, we analyzed the results of the
KSS bound and the WGC in section \ref{sec:5} and our correction terms shown that the joint
range between the two conjectures for the hyperscaling violation charged AdS$_{d+2}$ black
brane in the state $z=1, d=5, \theta=4.5$. The quantities must be measured at one temperature
for a meaningful comparison because these values are in different theories. We rewrite these
values by putting $\tilde{r}$ (expressed in (\ref{eq:3.21})) in them.
One of our most important results was that the correction signs in KSS bound (\ref{eq:4.8})
and WGC (\ref{eq:3.24}) are opposite (except term $\beta_2$). It means that the constraints
 are exclusive, and these theories may not be UV completed, i.e., four-derivative, Gauss-Bonnet,
 and eight-derivative. (Notice: this result is obtained for our particular black brane in extremal
 conditions, it is interesting to study these quantities for such a particular black brane.
 So there is an opportunity for possible changes in other conditions.) On the other hand,
 for the six-derivative coefficient $\beta_2$, both conjectures need the same sign.
 As a result, in the sense KSS bound and WGC, $\beta_2$ shown good behavior, which can
  mean that the ratio of $\frac{\eta}{s}$ does not vanish in extremal conditions.
  These results are consistent with the results of other black branes \cite{57,96},
  which could mean the universal behavior of corrections in these conjectures.
  We also considered first-order $\alpha'$ corrections in our study. It is possible
  that at higher-levels these two conjectures can be satisfied more beautifully and more widely.
In our future work, it may be interesting to consider  Maxwell field with other higher order derivative corrections.
These can serve as probes for the relationship between conjectures of the KSS bound and the WGC.
Also, we will rewrite our results in terms of parameters CFT, thereby expanding our analysis of these conjectures.
\appendix
\section*{Appendix}
\section{The Linearization}
\label{app:1}
The Riemann tensor, Ricci tensor, and the scalar curve in $d+2$-dimensions, used in equations (\ref{eq:3.9})
and (\ref{eq:3.10}).
\begin{equation}\label{eq:1A}
{\bar{R}}_{acbd}=\frac{2{\Lambda}_{eff}}{d(d+1)}(\bar{g}_{ab}\bar{g}_{cd}-\bar{g}_{ad}\bar{g}_{cb})\,,
\qquad{\bar{R}}_{ab}=\frac{2{\Lambda}_{eff}}{d}\bar{g}_{ab}\,,\qquad\bar{R}=\frac{2(d+2){\Lambda}_{eff}}{d},
\end{equation}
We present the linear equation of motion used in section \ref{subsec:3.1}.
 These calculations are in $d+2$-dimensions.

\begin{equation}\label{eq:2A}
\begin{split}
&{R}^2_{ab}{\big|}_{L}=\frac{4\Lambda}{d}{R}_L,\hspace{154pt} R{R}_{ab}{\big|}_{L}=\frac{2\Lambda}{d}((d+2){R}^L_{ab}+{\bar{g}}_{ab}{R}_L),\\
&{R}^L_{acbd}{\bar{g}}^{cd}{\big|}_{L}=\frac{2\Lambda}{d(d+1)}({\bar{g}}_{ab}h-{h}_{ab})+{R}^L_{ab},\hspace{34pt}{{R}^d}_a{R}_{bd}{\big|}_{L}=\frac{4\Lambda}{d^2}(d{R}^L_{ab}-\Lambda{h}_{ab}),\\
&{R}_{acbd}{R}^{cd}{\big|}_{L}=\frac{2\Lambda}{d+1}({R}^L_{ab}+\frac{{\bar{g}}_{ab}}{d}{R}_L+\frac{2\Lambda}{d^2}{h}_{ab}),\hspace{18pt}{R}_{abcd}{R}^{abcd}{\big|}_{L}=\frac{8\Lambda}{d(d+1)}{R}_L,\\
&{R}_{aecd}{{R}_{b}}^{ecd}{\big|}_{L}=\frac{8\Lambda}{d(d+1)}({R}^L_{ab}-\frac{\Lambda}{d}{h}_{ab}),\hspace{46pt}{R}^2_{abcd}-4{R}^2_{cd}+{R}^2{\big|}_{L}=\frac{4\Lambda(d-1)}{(d+1)}{R}_L,
\end{split}
\end{equation}
\begin{equation}\label{eq:3A}
\begin{split}
\nabla_f\nabla_c({{R}_a}^{qpf}{{R}^c}_{qpb}){\big|}_{L}=&\frac{2\Lambda}{d(d+1)}{\bar{\nabla}}_a{\bar{\nabla}}_b{R}_L+\frac{4\Lambda^2(d+2)}{d^2(d+1)^2}(({R}^L_{ab}-{R}_{L}\frac{{\bar{g}}_{ab}}{(d+2)})-\frac{2\Lambda}{d}{h}_{ab}),\\
\nabla_f\nabla_c({{R}_a}^{cpq}{{{R}_{pq}}^f}_b){\big|}_{L}=&\frac{-8\Lambda}{d(d+1)}(\bar{\Box}{R}^L_{ab}+\frac{1}{2}{\bar{\nabla}}_a{\bar{\nabla}}_b{R}_L)+\frac{16\Lambda^2}{d^2(d+1)^2}((d+1)\bar{\Box}{h}_{ab}-{R}_{L}{\bar{g}}_{ab}\\
&-(d+2){R}^L_{ab})-\frac{16(d+2)\Lambda^3}{d^3(d+1)^2}{h}_{ab},\\
\nabla_f\nabla_c({R}_{aqpb}{R}^{cqpf}){\big|}_{L}=&\frac{-2\Lambda}{d(1+d)}({\bar{\nabla}}_a{\bar{\nabla}}_b{R}_L-\bar{\Box}({R}^L_{ab}-\frac{{\bar{g}}_{ab}}{2}{R}_L))+\frac{8\Lambda^2}{d^2(1+d)^2}((2+d){R}^L_{ab}\\
&+\frac{(1+d)}{2}\bar{\Box}{h}_{ab}-{R}_{L}{\bar{g}}_{ab})-\frac{16(2+d)\Lambda^3}{d^3(1+d)^2}{h}_{ab},\\
\end{split}
\end{equation}
\begin{equation}\label{eq:4A}
\begin{split}
{R}_{dapc}{R}^{pcmn}{{R}^{d}}_{bmn}{\big|}_{L}=&\frac{24\Lambda^2}{d^2(d+1)^2}({R}^L_{ab}-\frac{4\Lambda}{3d}{h}_{ab}),\\
{{R}^n}_{acd}{{R}^{cq}}_{pn}{{{{R}^d}_q}^p}_b{\big|}_{L}=&\frac{8\Lambda^2}{d^2(d+1)^2}((d-1){R}^L_{ab}+\frac{{\bar{g}}_{ab}}{d}{R}_{L}-\frac{(d-2)\Lambda}{d}{h}_{ab}),\\
\end{split}
\end{equation}
\section{The Ansatz Solutions}\label{app:2}
The solution is calculated by placing the gravitational mass (\ref{eq:3.14}) in principal action (\ref{eq:2.1}).
 In fact, according to the solution non-corrections (\ref{eq:3.16}) and the cosmological constant (\ref{eq:3.10}),
 we compute the perturbation equations of motion up to the first order for each action term.
 This does, corrections the mass in the metric.
\begin{equation}\label{eq:1B}
\begin{split}
g_{tt}=&\big{[}-r^{4+2z-\theta}+Mr^{2-2z+\theta+\frac{\theta}{d}-d}-Q^2r^{2(3+\theta-\frac{\theta}{d}-d)}\big{]}+\alpha_1\big{[}2A_j-(d^2+2+3d)A_i\big{]}-\alpha_2\\
&\big{[}(d^2-3d+2)A_i\big{]}+\alpha_3\big{[}A_k+A_j-(1+d)A_i\big{]}-12\beta_1\big{[}2A_k+A_j+(d-\frac{3}{2})A_i\big{]}+3\beta_2\\
&\big{[}A_k+(\frac{3d^2}{2}-d)A_i\big{]}+(\gamma_1+2\gamma_3)\big{[}16A_j-(2+3d+d^2)A_i\big{]}+2\gamma_2\big{[}16A_j+(2+d)A_i\big{]}
\end{split}
\end{equation}
\begin{equation}\label{eq:2B}
\begin{split}
{g}_{rr}=&\big{[}\frac{{r}^{-2(\frac{\theta}{d}+3)}}{1-M{r}^{{\theta}-d-z}+{Q}^{2}{r}^{2({\theta}+1-d-z)}}\big{]}+{\alpha}_{1}\big{[}2{B}_{j}-(d^2+3d+2){B}_{i}\big{]}-{\alpha}_{2}\big{[}(d^2-3d+2)\\
&{B}_{i}\big{]}+{\alpha}_{3}\big{[}{B}_{k}+{B}_{j}-(d+1){B}_{i}\big{]}-12{\beta}_{1}\big{[}2{B}_{k}+{B}_{j}+(d-\frac{3}{2}){B}_{i}\big{]}+3{\beta}_{2}\big{[}(\frac{3d^2}{2}-d)\\
&{B}_{i}+{B}_{k}\big{]}+2({\gamma}_{1}+{\gamma}_{3})\big{[}16{B}_{j}-(d^2+3d+2){B}_{i}\big{]}+2{\gamma}_{2}\big{[}16{B}_{j}+(2+d){B}_{i}\big{]}
\end{split}
\end{equation}
and,
\begin{equation}\label{eq:3B}
\begin{split}
g_{xx}=&r^{2(1-\frac{\theta}{d})}+\alpha_1\big{[}2C_j-(d^2+3d+2)C_i\big{]}-\alpha_2\big{[}(d^2-3d+2)C_i\big{]}+\alpha_3\big{[}C_k+C_j-(1+d)\\
&C_i\big{]}-12\beta_1\big{[}2C_k+C_j+(d-\frac{3}{2})C_i\big{]}+3\beta_2\big{[}C_k+d(\frac{3d}{2}-1)C_i\big{]}+2(\gamma_1+2\gamma_3)\big{[}16C_j-\\
&(d^2+3d+2)C_i\big{]}+2\gamma_2\big{[}16C_j+(2+d)C_i\big{]}
\end{split}
\end{equation}
We have summarized the above equations and is written by.
\begin{equation}\label{eq:4B}
\begin{split}
&R_{xx}=r^2(1-\frac{\theta}{d})\big{[}\theta-z-\frac{1}{d}+Q^2\frac{r^{2(\theta+1)}}{r^{2(z+d)}}(z-\theta-d)\big{]},\\
&R_{tt}=\big{[}r^{2z}-M\frac{r^{z+\theta}}{r^{d}}+Q^2\frac{r^{2(1+\theta)}}{r^{2d}}\big{]}\big{[}(d-\theta+z)(z-\frac{\theta}{d})+Q^2\frac{r^{2(1+\theta+z)}}{r^{2d}}(z-\theta+d-2)\\
&(d-1-\theta+\frac{\theta}{d})\big{]},\\
&R_{rr}=\frac{(\theta-d)r^{z+d-2}\big{[}1+\frac{z}{d}+z\frac{(z-1)}{(d-\theta)}-\frac{Mr^{\theta}}{r^{z+d}}(\frac{\theta}{d}-z+1)+Q^2\frac{r^{2(\theta+1)}}{r^{2(z+d)}}(\frac{(2\theta-z+2)}{d}-\theta+d-2)\big{]}}{r^{-z-d}-Mr^{\theta}+Q^2r^{2\theta}},\\
&R=r^{\frac{2\theta}{d}}\big{[}(2z-\theta)\frac{\theta}{d}-d(2z+d+1)-2z^2+2\theta(z-\theta+d+1)\big{]}-M\frac{r^{\theta+\frac{2\theta}{d}}}{r^{z+d}}\big{[}d-z\theta+\frac{\theta^3}{d}\\
&+dz\big{]}+Q^2\frac{r^{2(1+\theta+\frac{\theta}{d})}}{r^{2(z+d)}}\big{[}2z-\theta(6+\theta)-4+d(3+2\theta-d)+(4+3\theta-2z)\frac{\theta}{d}\big{]},
\end{split}
\end{equation}
\begin{equation}\label{eq:5B}
\begin{split}
\Upsilon_i=&R_{\iota\iota}+g_{\iota\iota}(d+1-\frac{R}{2}),\hspace{6pt}\Upsilon_j=(d+1)(g_{\iota\iota}R+\frac{R''}{g_{\iota\iota}''}),\hspace{6pt}\Upsilon_k=g_{\iota\iota}((d+1)R+\frac{\Upsilon_i''}{g_{\iota\iota}''}),\\
&\Upsilon (g_{\iota\iota})\rightarrow A(g_{tt}),B(g_{rr}),C(g_{xx}),
\end{split}
\end{equation}
\section{The Hawking Temperatur}
\label{app:3}
By placing correction solutions \ref{app:2} in the Hawking temperature equation (\ref{eq:3.17}),
we obtain the temperature of the hyperscaling charged AdS$_{d+2}$ black brane. We consider $d=5$, $z=1$
and we get an average of $\theta=4.5$ from equation $d+z-2\leq\theta\leq d+z-1$. We also consider $r=r_h$ and
$M=r^{1/2}(1+Q^2r)$ by inserting the given values,
\begin{equation}\label{eq:1C}
\begin{split}
T=&\frac{1}{4\pi}(1.5r^5+0.5Q^2r^4+{\alpha}_{1}A_1+{\alpha}_{2}A_2+{\alpha}_{3}A_3+{\beta}_{1}B_1+{\beta}_{2}B_2+({\gamma}_{1}+{\gamma}_{3})C_1+{\gamma}_{2}C_2),
\end{split}
\end{equation}
\begin{equation}\label{eq:2C}
\begin{split}
A_1=&\big{[}0.1Q^4r^{\frac{-12}{5}}-2.3Q^8+2.9Q^2r^{\frac{-7}{5}}+12.5r^{\frac{-2}{5}}-21.8Q^6r-62.4Q^4r^2-31.9Q^2r^3-\\
&73Q^6r^{\frac{16}{5}}+69.7r^4+0.1Q^8r^4-750.5Q^4r^{\frac{21}{5}}+1.4Q^6r^5-2550.2Q^2r^{\frac{26}{5}}-35.1Q^4r^6-\\
&2873.2r^{\frac{31}{5}}-203.5Q^2r^7-272.4r^8\big{]}\big{/}\big{[}(0.1Q^4r+0.7Q^2r^2+1.2r^3)((-0.05Q^2-2.6r\\
&-1.1Q^4r^{\frac{17}{5}}-1.1Q^6r^{\frac{32}{5}}-1.1Q^2r^{\frac{33}{5}}-1.08Q^4r^{\frac{37}{5}}-1.1Q^2r^{\frac{42}{5}})10^{16}+47.2Q^2r^{\frac{22}{5}}+\\
&63r^{\frac{27}{5}}+552r^{\frac{47}{5}})^{\frac{1}{2}}\big{]},\\
A_2=&0,
\end{split}
\end{equation}
\begin{equation}\label{eq:3C}
\begin{split}
A_3=&r^{\frac{41}{10}}\big{[}0.07Q^8-0.05Q^4r^{\frac{-12}{5}}-1.7Q^2r^{\frac{-7}{5}}-7.5r^{\frac{-2}{5}}+0.6Q^6r+1.9Q^4r^2+1.01r^3Q^2+\\
&1.7Q^6r^{\frac{16}{5}}-1.9r^4-0.1Q^8r^4+17.8Q^4r^{\frac{21}{5}}-0.4Q^6r^5+60.5Q^2r^{\frac{26}{5}}+4.3Q^4r^6+24.8\\
&Q^2r^7+68.2r^{\frac{31}{5}}+33.6r^8\big{]}\big{/}\big{[}(0.2Q^4+2.05Q^2r+3.7r^2)i(1+48.4rQ^{-2})^{\frac{1}{2}}\big{]},
\end{split}
\end{equation}
\begin{equation}\label{eq:4C}
\begin{split}
B_1=&r^{\frac{51}{10}}\big{[}16.5r^4-0.6Q^4r^{\frac{-12}{5}}-21.6Q^2r^{\frac{-7}{5}}-90.8r^{\frac{-2}{5}}-3.9Q^6r-11.4Q^4r^2-3.9r^3Q^2-\\
&23.1Q^6r^{\frac{16}{5}}-0.4Q^8-1.6Q^8r^4-237.03Q^4r^{\frac{21}{5}}-6.2Q^6r^5-804.9Q^2r^{\frac{26}{5}}+51.2r^6Q^4-\\
&907.5r^{\frac{31}{5}}+304.1Q^2r^7+416.4r^8\big{]}\big{/}\big{[}(Q^4r+7.2Q^2r^2+13.04r^3)i(1+48.4rQ^{-2})^{\frac{1}{2}}\big{]},
\end{split}
\end{equation}
\begin{equation}\label{eq:5C}
\begin{split}
B_2=&Qr^{\frac{12}{5}}\big{[}4.8Q^8+45.8Q^6r+130.9Q^4r^2+62.5Q^2r^3+150.7Q^6r^{\frac{16}{5}}-144.2r^4-1.1r^4Q^8+\\
&1541.3Q^4r^{\frac{21}{5}}-3.3Q^6r^5+5234.3Q^2r^{\frac{26}{5}}+48.8Q^4r^6+5901.09r^{\frac{31}{5}}+259.1Q^2r^7+r^8\\
&344.1\big{]}\big{/}\big{[}(\sqrt{2}Q^4+10.2Q^2r+18.4r^2)(3Q^4r^4-2Q^4+64Q^2r^{\frac{16}{5}}+256r^{\frac{21}{5}}-32r^6)^{\frac{1}{2}}\big{]},
\end{split}
\end{equation}
\begin{equation}\label{eq:6C}
\begin{split}
C_1=&r^{\frac{46}{10}}\big{[}0.03Q^4r^{\frac{-12}{5}}-0.1Q^8+1.1Q^2r^{\frac{-7}{5}}+4.7r^{\frac{-2}{5}}-1.03Q^6r-2.9Q^4r^2-1.5Q^2r^3-3.4\\
&Q^6r^{\frac{16}{5}}+3.3r^4+0.03Q^8r^4-35.6Q^4r^{\frac{21}{5}}+0.19Q^6r^5-121.02Q^2r^{\frac{26}{5}}-2.6Q^4r^6-136.3\\
&r^{\frac{31}{5}}-15.6Q^2r^7-20.9r^8\big{]}\big{/}\big{[}(0.1Q^2r+0.7r^2+1.2r^3Q^{-2})(0.05Q^2r^{\frac{22}{5}}-4.8\times10^{12}Q^2\\
&-2.3\times10^{14}r-1.27\times10^{13}Q^4r^{\frac{17}{5}}+0.07r^{\frac{27}{5}}-1.27\times10^{13}Q^6r^{\frac{32}{5}}-1.27\times10^{13}r^{\frac{33}{5}}Q^2\\
&-7.9\times10^{12}Q^4r^{\frac{37}{5}}-1.27\times10^{13}Q^2r^{\frac{42}{5}}+r^{\frac{47}{5}})^{\frac{1}{2}}\big{]},
\end{split}
\end{equation}
\begin{equation}\label{eq:7C}
\begin{split}
C_2=&r^{\frac{46}{10}}\big{[}0.03Q^8+0.06Q^4r^{\frac{-12}{5}}+1.9Q^2r^{\frac{-7}{5}}+8.19r^{\frac{-2}{5}}+0.29Q^6r+0.84Q^4r^2+0.4Q^2r^3+\\
&0.99Q^6r^{\frac{16}{5}}-0.9r^4+0.06Q^8r^4+10.2Q^4r^{\frac{21}{5}}+0.2 Q^6r^5+34.7Q^2r^{\frac{26}{5}}-1.5Q^4r^6+r^{\frac{31}{5}}\\
&39.09-9.19Q^2r^7-12.3r^8\big{]}\big{/}\big{[}(-0.1Q^2r-0.7r^2-1.2r^3Q^{-2})(-1.4\times10^{13}Q^2-7.06\\
&\times10^{14}r+6.3\times10^{12}Q^4r^{\frac{17}{5}}-0.02Q^2r^{\frac{22}{5}}-0.03r^{\frac{27}{5}}+6.3\times10^{12}Q^6r^{\frac{32}{5}}+6.3\times10^{12}\\
&r^{\frac{33}{5}}Q^2+2.07\times10^{13}Q^4r^{\frac{37}{5}}+6.3\times10^{12}Q^2r^{\frac{42}{5}}+r^{\frac{47}{5}})^{\frac{1}{2}}\big{]},
\end{split}
\end{equation}
\section{The Ratio $\eta/s$}
\label{app:4}
We obtain the ratio $\frac{\eta}{s}=TD$ using equations appendix \ref{app:2}, \ref{app:3}, and (\ref{eq:4.6})
for $Q^2=-3r$ and $M=r^{1/2}(1+Q^2r)$, $r=r_h$.
\begin{equation}\label{eq:1D}
\begin{split}
\frac{\eta}{s}=\frac{1}{4\pi}(-1.36-0.45Q^2+\alpha_1A_1+\alpha_2A_2+\alpha_3A_3+\beta_1B_1+\beta_2B_2+\gamma_1C_1+\gamma_3C_1+\gamma_2C_2)
\end{split}
\end{equation}
\begin{equation}\label{eq:2D}
\begin{split}
A_1=&\big{[}(0.149-1.19Q^2+19.07Q^4+1.19Q^{32}-0.018Q^{36}+0.0005Q^{38}-0.000009Q^{40})\\
&10^{-7}+0.00003Q^8-0.00003Q^{10}+0.0002Q^{16}-0.0001Q^{18}-0.0002Q^{20}-0.00001\\
&Q^{26}\big{]}\big{/}\big{[}74.08+1069.3Q^2+7263Q^4+30894Q^6+92400.7Q^8+206870Q^{10}+360336\\
&Q^{12}+500836Q^{14}+564680.6Q^{16}+521348Q^{18}+395338.9Q^{20}+245366.7Q^{22}+Q^{24}\\
&123362+49346.06Q^{26}+15277.4Q^{28}+3511.9Q^{30}+561Q^{32}+55.3Q^{34}+2.5Q^{36}\big{]},\\
A_2=&0,
\end{split}
\end{equation}
\begin{equation}\label{eq:3D}
\begin{split}
A_3=&\big{[}1.16+18.6Q^2+74.5Q^4-596.04Q^8-2384.1Q^{10}-4768.3Q^{12}-2384.1Q^{14}-4768\\
&Q^{16}-2384.1Q^{18}-149.01Q^{26}-37.2Q^{28}+0.03Q^{34}+0.002Q^{36}+7\times10^{-5}Q^{38}+2.2\\
&\times10^{-6}Q^{40}\big{]}10^{-10}\big{/}\big[38.4+521.9Q^2+3289.9Q^4+12764.5Q^6+34075.6Q^8+66253.8\\
&Q^{10}+96848.1Q^{12}+108285.1Q^{14}+93275.7Q^{16}+61832Q^{18}+31250.3Q^{20}+11814.8\\
&Q^{22}+3234.3Q^{24}+607.05Q^{26}+70.8Q^{28}+4.17Q^{30}+0.06Q^{32}-1.1\times10^{-13}Q^{34}\big],
\end{split}
\end{equation}
\begin{equation}\label{eq:4D}
\begin{split}
B_1=&\big{[}10^{-8}(5.9Q^2-71.5Q^4-572.2Q^8+762.9Q^{10}-47.6Q^{26}-2.98Q^{28}+0.01Q^{36}-\\
&0.001Q^{38})-0.00003Q^{12}+0.00003Q^{14}-0.00001Q^{16}+0.00003Q^{18}\big{]}\big{/}\big{[}461.4+Q^2\\
&6263+39478Q^4+153174Q^6+408908Q^8+795045.9Q^{10}+1162177Q^{12}+1299422\\
&Q^{14}+1119309Q^{16}+741984Q^{18}+375004.5Q^{20}+141778Q^{22}+38812.5Q^{24}+Q^{26}\\
&7284.6+850.15Q^{28}+50.1Q^{30}+0.7Q^{32}+1.36\times10^{-12}Q^{34}+7.1\times10^{-14}Q^{36}\big{]},
\end{split}
\end{equation}
\begin{equation}\label{eq:5D}
\begin{split}
B_2=&\big{[}1862.6+14901Q^2+238418.5Q^4+357627.8Q^6+238418.5Q^8-476837Q^{10}-Q^{12}\\
&953674.3+476837Q^{14}+476837Q^{16}+119209.2Q^{20}+7450.5Q^{22}+1862.6Q^{24}+\\
&Q^{28}0.01+0.05Q^{30}\big{]}\big{/}\big{[}698.4Q^6-1.81-291.03Q^4+931.3Q^8+465.6Q^{12}-465.6\\
&Q^{14}-232.8Q^{16}-58.2Q^{18}-29.1Q^{20}+3.6Q^{22}+0.02Q^{24}-0.007Q^{26}\big{]},
\end{split}
\end{equation}
\begin{equation}\label{eq:6D}
\begin{split}
C_1=&\big{[}(76.2Q^2-4.7+38.1Q^{30}-9.5Q^{32}-1.1Q^{34}-0.1Q^{36}-0.002Q^{38})10^{-7}+0.00003\\
&Q^4-0.0002Q^8+0.0009Q^{12}+0.001Q^{16}-0.0009Q^{18}+0.0009Q^{20}-0.0001Q^{26}-\\
&0.00003Q^{28}\big{]}\big{/}\big{[}1185.4+16220.06Q^2+103269.5Q^4+405972.2Q^6+778772.5Q^{22}+\\
&318289.9Q^{24}+109098Q^{26}+30833.5Q^{28}+6834.7Q^{30}+1090.4Q^{32}+108.9Q^{34}+\\
&5.04Q^{36}+(1.1Q^8+2.1Q^{10}+3.3Q^{12}+3.8Q^{14}+3.6Q^{16}+2.65Q^{18}+1.5Q^{20})10^6\big{]},
\end{split}
\end{equation}
\begin{equation}\label{eq:7D}
\begin{split}
C_2=&\big{[}((4.7Q^2+19Q^4+38Q^{26}+1.1Q^{30}-0.14Q^{32})10^5-174.6Q^{34}-58.2Q^{36}-3.6Q^{38}\\
&)10^{-12}+0.00002Q^6+0.00003Q^8-0.0001Q^{10}-0.00006Q^{12}+0.0001Q^{14}+0.0001\\
&Q^{16}-0.0001Q^{18}-0.0001Q^{20}-0.00001Q^{24}\big{]}\big{/}\big{[}-1185.4-16071.8Q^2-101112Q^4\\
&-391249.8Q^6-798581.7Q^{20}-254256.5Q^{22}-42372Q^{24}+4307.6Q^{26}+4767.5Q^{28}\\
&+1391.3Q^{30}+224.3Q^{32}+20.3Q^{34}+0.8Q^{36}+(-1.04Q^8-2.01Q^{10}-Q^{12}2.9-\\
&3.2Q^{14}-2.6Q^{16}-1.7Q^{18})10^6\big{]},
\end{split}
\end{equation}

\end{document}